\documentclass[aps,twocolumn,prb,amsmath,amssymb]{revtex4-1}
\usepackage[varg]{txfonts}
\usepackage{color}
\usepackage[colorlinks]{hyperref}

\usepackage{graphicx}
\usepackage{bm}

\DeclareMathOperator{\Ree}{Re}

\def\XXint#1#2#3{{\setbox0=\hbox{$#1{#2#3}{\int}$}
    \vcenter{\hbox{$#2#3$}}\kern-.5\wd0}}

\def\be{\begin{equation}}
\def\ee{\end{equation}}

\def\bi{\begin{itemize}}
    \def\ei{\end{itemize}}
\def\bn{\begin{enumerate}}
    \def\en{\end{enumerate}}
\def\bea{\begin{eqnarray}}
\def\eea{\end{eqnarray}}
\newcommand{\bpm}{\begin{pmatrix}}
    \newcommand{\epm}{\end{pmatrix}}

\def\ba{\begin{array}}
    \def\ea{\end{array}}
\def\bd{\begin{displaymath}}
\def\ed{\end{displaymath}}

\renewcommand{\imath}{\hspace{1pt}\mathrm{i}\hspace{1pt}}

\renewcommand{\Re}{\mathop{\mathrm{Re}}\nolimits}
\renewcommand{\Im}{\mathop{\mathrm{Im}}\nolimits} 

\begin{document}

\title{Excitonic insulator phase and dynamics of condensate in a topological one-dimensional model}

\author{Zahra Khatibi}
\affiliation{Department of Physics, Sharif University of Technology, Tehran 14588-89694, Iran}

\author{Roya Ahemeh}
\affiliation{Department of Physics, Sharif University of Technology, Tehran 14588-89694, Iran}

\author{Mehdi Kargarian}
\email{kargarian@physics.sharif.edu}
\affiliation{Department of Physics, Sharif University of Technology, Tehran 14588-89694, Iran}

\begin{abstract}
We employ mean-field approximation to investigate the interplay between the nontrivial band topology and the formation of excitonic insulator (EI) in a one-dimensional chain of atomic $s-p$ orbitals in the presence of repulsive inter-orbital Coulomb interaction. We find that our model, in a non-interacting regime, admits topological and trivial insulator phases, whereas, in strong Coulomb interaction limit, the chiral symmetry is broken and the system undergoes a topological-excitonic insulator phase transition. The latter phase transition stems from an orbital pseudomagnetization and band inversion around $k=0$. Our findings show that contrary to the topological insulator phase, electron-hole bound states do not form exciton condensate in the trivial band insulator phase due to lack of band inversion. Interestingly, the EI phase in low $s-p$ hybridization limit hosts a Bardeen-Cooper-Schrieffer (BCS)/Bose-Einstein condensation (BEC) crossover. Irradiated by a pump pulse, our findings reveal that the oscillations of exciton states strongly depend on the frequency of the laser pulse. We further explore the signatures of dynamics of the exciton condensate in optical measurements. 
\end{abstract}

\maketitle

\section{Introduction}\label{Introduction}
The many-body problem of exciton formation driven by charge instability and Coulomb attraction between electron-hole pairs has triggered intensive previous and contemporary investigations in bulk and low dimensional semiconductors \cite{Salvo1986physical,Wakisaka2009,hellmann2012time,Zenker2013chiral,Kaneko2015,Larkin2017Giant,Remez2020,Inayoshi2020,kadosawa2020finite,murakami2020ultrafast}. 
Excitonic insulators (EI) arising from the condensation of electron-hole bound states, despite being conceptually introduced decades ago \cite{jerome1967excitonic,kohn1967excitonic,halperin1968possible}, have received new attention in recent years due to possible realization in the bulk of semiconductors \cite{Seki2014Excitonic,kim2016layer}.

Few layered transition metal dichalcogenide (TMD), are reputedly known to host collective exciton condensation due to low dimensionality and strong light-matter coupling \cite{Yu2014Dirac,Zhou2015berry,Srivastava2015}. The prominent two-dimensional devices based on strong moire periodic potential in nearly aligned heterostructures of TMDs facilitate the observation of band dispersion flattening with the formation of strongly bound states. These devices are interesting platforms for various excitonic states studies, such as topological exciton bands and strongly correlated exciton Hubbard model \cite{jin2019observation}. In nearly aligned WSe$_2$/WS$_2$, within the A-exciton spectra region of WSe$_2$ layer, the spatially localized interlayer excitons have been reported to respond differently to back gate doping. The reason is ascribed to the spatial distribution of exciton wave function and electron-exciton interactions based on the electron doping region \cite{jin2019observation}.
Thick encapsulation of TMDs with hexagonal Boron Nitride (hBN) layers leading to a weak coupling regime at nearly zero temperature can tune the coherent radiative decay rates of neutral excitons in these materials. Spontaneous photo-induced radiative recombination is believed to result in a photoluminescence spectra rise during the laser pump exposure, while the high-energy exciton relaxation to radiative states dominates the post-laser pulse exposure. In fact, the suppression of environmental dielectric constant through the hBN encapsulation results in radiative lifetime enhancement \cite{fang2019control} and suppression of nonradiative processes due to nearly zero temperature and the least existence of disorder and contamination \cite{cadiz2017excitonic}.

Another intriguing platform to study the spontaneous exciton condensation is the promising quasi-one-dimensional (1D) chalcogenide Ta$_2$NiSe$_5$ \cite{lu2017zero,Werdehausen2018,Mazza2019,Tang2020,andrich2020imaging}. The large bandgap opening fingerprint in photoemission spectroscopy in a recent study, is believed to mark the enhancement of exciton order in the spatially separated Ni and Ta chains \cite{Werdehausen2018}. Moreover, the newly reported novel low-frequency mode in Raman spectra was proposed as evidence for the existence of an EI phase in Ta$_2$NiSe$_5$ emerging below $\approx$328~K \cite{Werdehausen2018}. Further analytical investigations suggested that the phase transition is associated with a Bose-Einstein condensation (BEC) in the scheme of a one-dimensional (1D) extended Falicov-Kimball model (EFKM) with an overlapping band semimetal as the normal state \cite{Sugimoto2018strong}. Meanwhile, another plausible scenario is argued to be a spontaneous Ta-Ni hybridization based on charge instability which breaks the symmetry \cite{Mazza2019}. Besides, the importance of structural phase transition as origin of electronic gap has been addressed in recent pump-probe \cite{Baldini2020spontaneous} and Raman spectroscopy \cite{Kim2020Observation} measurements.

To date, the theorized models to understand the EI phase are mostly based on the idea of strong correlations in a semimetal with a small band overlap or a semiconductor with a small gap \cite{jerome1967excitonic,Zenker2010,hellmann2012time}. Given the importance of designing new models, here, we explore the EI phase in a 1D chain of atomic $s-p$ orbitals with the inclusion of odd parity hybridization \cite{shockley1939surface,continentino2014topological} (See Fig.~\ref{scheme}(a)). In the non-interacting scheme, this model exhibits topological insulator (TI) and trivial band insulator (BI) phases \cite{continentino2014topological}. Adding the inter-orbital Coulomb interaction, our main objection is to answer the following questions. How do the latter phases change with correlations? Do the TI and EI phases compete or coexist? In the condensed phase, how do the collective modes affect the optical transitions? 
How does the non-equilibrium dynamics of exciton condensate, coupled to a phonon bath, respond to ultrafast pulses, and what are the experimental consequences in optical measurements? 

In this article, we show that the EI phase emerges out of the topological insulator phase beyond a critical interaction, while the BI phase remains remote in forming exciton condensate. We show that both amplitude and phase collective modes  are  gaped which are manifest as many-body excitations in optical responses.
Furthermore, our non-equilibrium analysis reveals that the oscillation of exciton condensate strongly depends on the frequency of driving pulse with signatures visible in optical conductivity.

The paper is organized as follows. In section \ref{model} we present the theoretical model for a 1D $s-p$ chain in equilibrium and address the nature of spontaneous exciton condensation in a topological insulator phase. The linear response and collective mode signatures are addressed in the last part of this section.
Next, we discuss the exciton dynamics in a stimulated pump-probe situation in section \ref{noneq}. In section \ref{opcosec}, we present the results for the optical spectra of a $s-p$ chain in a linear response regime. Our findings are summarized in section \ref{concolusion}.

         
\section{model and equilibrium phase diagram}\label{model}
In this section, we introduce the 1D $s-p$ chain and present a comprehensive analysis of its equilibrium phase diagram.

\subsection{Interacting s-p model }
\begin{figure}[t]
    \center
    \includegraphics[width=\linewidth]{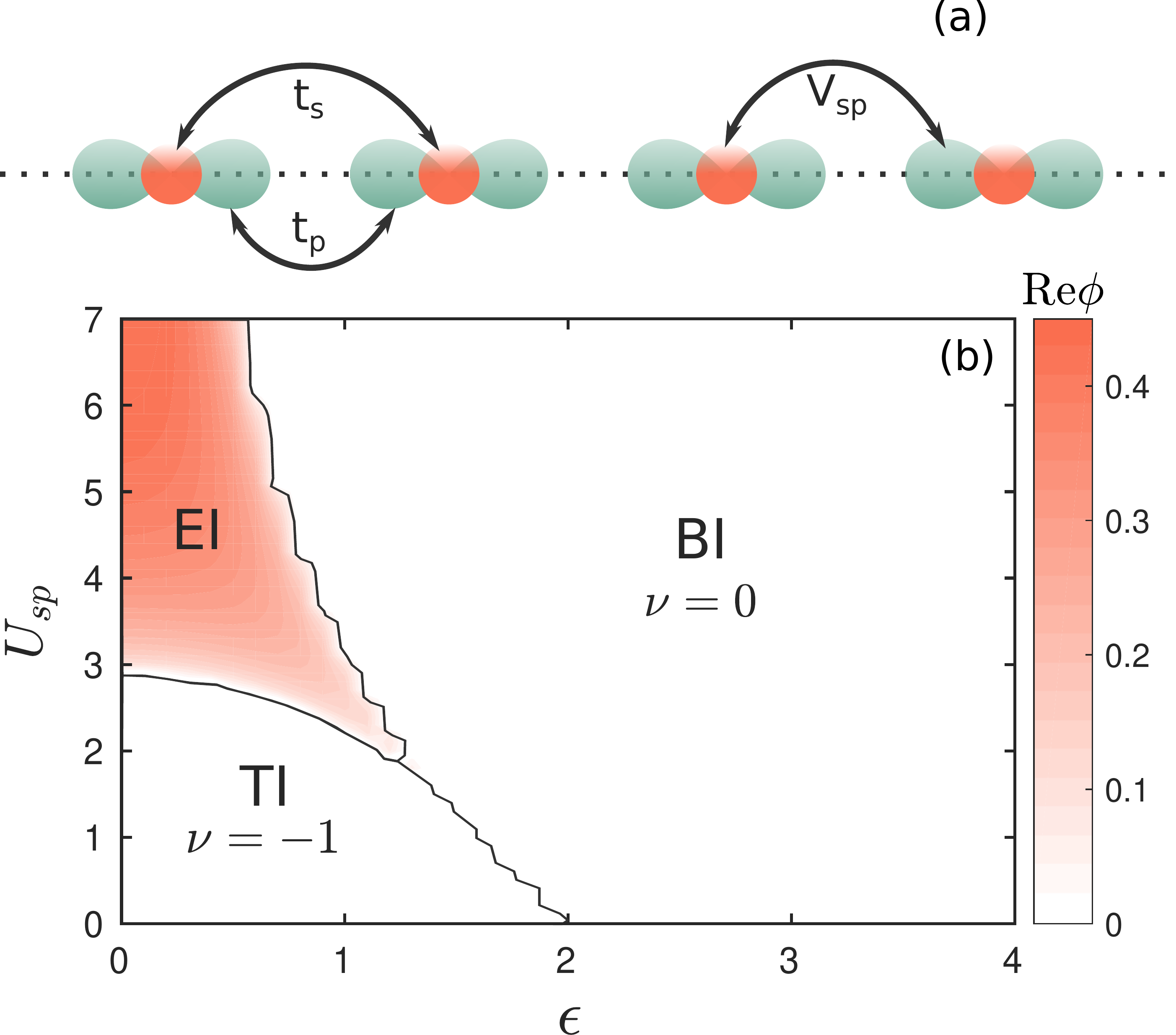}
    \caption{(a) Schematic illustration of a 1D $s-p$ chain. The sphere-shaped $s$ orbitals and dumbbell-shaped $p$ orbitals are marked in orange and green colors, respectively. $t_s(t_p)$ and $V_{sp}$ are the intra and inter-orbital tunneling parameters. (b) Contour plot of exciton parameter real part as a function of on-site energy ($\epsilon$) and Coulomb interaction ($U_{sp}$) for $V_{sp}=1/2$. The TI phase with winding number $\nu=-1$ and BI phase, share the zero exciton region (white area). In a strongly coupled $s-p$ chain, on the other hand, the chiral symmetry is broken and the system undergoes a phase transition from TI to EI where excitons emerge. \label{scheme}}
\end{figure}

The non-interacting spinless model of a 1D chain of atoms with $s$ and $p$ orbitals and lattice spacing $a$, as shown in Fig.~\ref{scheme}(a), reads \cite{shockley1939surface,continentino2014topological,kunevs2015excitonic},
\begin{align}
\label{sp}
H_0&=\epsilon_s \sum_j c^{\dagger}_j c_j + \epsilon_p \sum_j p^{\dagger}_j p_j- 
\sum_j t_s (c^{\dagger}_j c_{j+1} + c^{\dagger}_{j+1} c_{j})\nonumber \\
&+ \sum_j t_p (p^{\dagger}_j p_{j+1} + p^{\dagger}_{j+1} p_{j}) 
+V_{sp} \sum_j (c^{\dagger}_j p_{j+1} -c^{\dagger}_{j+1} p_{j}) \nonumber \\
&- V_{ps} \sum_j (p^{\dagger}_j c_{j+1} - p^{\dagger}_{j+1} c_{j})  
\end{align}
where $c^{\dagger}_j (c_j)$ and $p^{\dagger}_j(p_j)$ are the charge creation and annihilation operators in $s$ and $p$ orbitals of $j$th  atomic site, respectively. $\epsilon_s(\epsilon_p)$ is the on-site energy, and $t_s(t_p)$ is the hopping parameter between nearest neighbors with same orbitals. Also, $V_{sp} (V_{ps})$ is the hybridization energy between $s(p)$ and $p(s)$ orbitals in a neighboring site with an odd parity, i.e. $V_{sp}(-x)=-V_{sp}(x)$. This odd parity is responsible for the band inversion that leads to TI phase formation \cite{continentino2014topological}. By Fourier transformation to momentum space, the Hamiltonian \eqref{sp} becomes

\begin{align}\label{nonintH}
H_0&=\sum_{k}\left(\epsilon_s-2t_s  \cos ka  \right)c^{\dagger}_{k} c_k +\sum_{k}\left( \epsilon_p +2t_p \cos ka \right)
\nonumber p^{\dagger}_{k} p_k\\
&+ 2i V_{sp}\sum_{k} \sin ka ~c^{\dagger}_{k} p_{k} - 2i V_{ps}\sum_{k} \sin ka ~p^{\dagger}_{k} c_{k}.
\end{align}

In a more generic model where repulsive short-range Coulomb interactions are present, the interacting Hamiltonian is $H_e=H_0+H_U$, where 

\begin{align}\label{Hu}
H_U=U_{sp} \sum_{j}(c^{\dagger}_j c_j-1/2)(p^{\dagger}_j p_j-1/2).
\end{align}
Here, $U_{sp}$ is the strength of inter-orbital Coulomb interaction between spinless electrons residing in local $s$ and $p$ orbitals. We treat the above interaction using the mean-field approximation and decouple the local two-body terms into density and exciton order parameter channels \cite{murakami2017photoinduced,murakami2020ultrafast}. Fourier transforming Eq.~\eqref{Hu}, we obtain

\begin{align}
U_{sp} \sum_k \left[ \left(n_s-1/2\right) p_{k}^\dagger p_{k} + \left(n_p-1/2\right) c_{k}^\dagger c_{k}\nonumber 
-\phi p_{k}^\dagger c_{k}- \phi^\ast c_{k}^\dagger p_{k}\right],
\end{align}
%
%
where, $\phi=\langle c_{j}^\dagger p_{j}\rangle$ is the exciton order parameter. Moreover, $n_s=\langle c^{\dagger}_j c_j\rangle$ and $n_p=\langle p^{\dagger}_j p_j\rangle$ are the charge density order parameters of $s$ and $p$ orbitals, respectively. In momentum space the two-band mean-field Hamiltonian is cast as a pseudospin in a pseudomagnetic field \cite{tanabe2018nonequilibrium}:

\begin{align}\label{b_spin}
H_e^{\rm MF}=\sum_{k,\gamma}B^{\gamma}_k S^{\gamma}_k,
\end{align}
where $S^{\gamma}_k$ is the pseudospin component,
\begin{align}\label{Hspinor}
S^{\gamma}_{k}=\frac{1}{2}\zeta^{\dagger}_{k}\sigma_{\gamma}\zeta_{k},~~\zeta^{\dagger}_k=\left(
c^\dagger_k,p^\dagger_k\right)
\end{align}
with $\sigma_{\gamma}~(\gamma=1-3)$ being the Pauli matrices, and $\sigma_0$, the identity matrix. In what follows, we set the parameters as $\epsilon_s=-\epsilon_p=\epsilon$, $t_s=t_p=t$ and $V_{sp}=V_{ps}$. Also, for sake of simplicity, we set $a=1$, and $t=1$, hereafter, unless otherwise is stated. With these identifications, the components of the pseudomagnetic field, $B^\gamma_k$s, are 
\begin{subequations}\label{reducedB}
\begin{align}
B^0_k&=U_{sp}(n_s+n_p-1) \\
B^1_k&=-2 U_{sp} \Re{\phi}\\
B^2_k&=-2 U_{sp} \Im{\phi}-4V_{sp}\sin k   \\
B^3_k&= 2\epsilon-4t \cos k  +U_{sp}(n_p-n_s).
\end{align} \end{subequations} 

Now, we adopt a self-consistent calculation to address the energy dispersion and the order parameters in equilibrium state at zero temperature. The procedure is as follows. First, we solve the eigenvalue problem $\left(H^{\rm MF}_{k}-E_{k, \pm}\right)|k,\pm\rangle=0$, for the Bloch Hamiltonian driven from Eq.~\eqref{Hspinor}, to find the eigenvalues, $E_{k, \pm}=\big[B^{0}_{k}\pm |{\bf B}_{k}|\big]/2$, and their corresponding eigen functions, $|k,\pm\rangle$. Next, we use this knowledge to evaluate  the expectation values of pseudospin components at equilibrium using the following relation \cite{tanabe2018nonequilibrium}

\begin{align}\label{expectvaluesp}
\left\langle S^\gamma_k(0) \right\rangle =
 \begin{cases}
\displaystyle{\frac{B^\gamma_k(0)}{2 \left| {\bf B}_k(0) \right|}} \left[ f (E_{k, +}(0)) - f (E_{k, -}(0))  \right]
& (\gamma = 1-3)
\\
\displaystyle{\frac{1}{2}} \left[f (E_{k, +} (0)) + f (E_{k, -} (0)) \right]
& (\gamma = 0),
\end{cases}
\end{align}
where $f (E_{k, +}(0))$ is the Fermi-Dirac distribution function at equilibrium. We start with an initial guess for $\left\langle S^\gamma_k(0) \right\rangle$ over all momentum vector $k$. We then obtain the mean-field order parameters via
\begin{align}\label{phi-n0-n1}
    \begin{bmatrix}
        n_0&\phi^\ast\\\phi &n_1
    \end{bmatrix}
    =\frac{1}{N}\sum_k\left[\langle {\bf S}_k\rangle \cdot {\boldsymbol \sigma}+\langle S^0_k\rangle\sigma_0\right].
\end{align}

Eventually, the latter values are fed into $H_{e}^{\mathrm{MF}}$ to find $\left\langle S^\gamma_k(0) \right\rangle$ through Eq.~\eqref{expectvaluesp}. Until reaching the convergence, this process is iterated with the following assumption that the system is at half filling, i.e., $n_s+n_p=1$. 

Before we proceed any further, it is worthwhile to note a few points about the symmetry in the EI phase. In the absence of $s-p$ orbitals hybridization, $V_{sp}$, in Hamiltonian \eqref{sp}, both valence and conduction bands enjoy $U(1)$ charge conservation symmetry separately. This implies that the overall symmetry is $U_s(1)\times U_p(1)$ for $s$ and $p$ bands, which is spontaneously broken in EI phase due to condensation of complex exciton order parameter, $\phi=|\phi|e^{i\varphi}$. Here, $\varphi$ is the phase of the condensate whose fluctuations give rise to the phase mode \cite{Remez2020,Murakami2020Collective}. The net symmetry in our model, however, is explicitly broken down to a $U(1)$ symmetry of total charge conservation by a nonzero $V_{sp}$ (and also by coupling to phonons which we will discuss in Sec.~\ref{phonon_coupling}). We further find that the order parameter $\phi$ becomes real, $\mathrm{Im}\phi=0$, at equilibrium, which could be attributed to locking of the order parameter to a particular direction due the aforementioned symmetry breaking in the equilibrium ground state.

\subsection{Equilibrium phase diagram\label{phase_equ}}
The equilibrium phase diagram in the plane of $U_{sp}$ and on-site energy for a fixed value of inter-orbital hybridization $V_{sp}=1/2$ is shown in Fig.~\ref{scheme}(b). In the non-interacting limit, i.e., $U_{sp}=0$, a TI phase sets in for $\epsilon<2$, while the trivial BI phase appears for $\epsilon>2$, consistent with previous studies \cite{continentino2014topological}. In the TI phase, the bands undergo an inversion around $k=0$ and thus the winding number becomes $\nu=-1$. In the BI phase, on the other hand, the valence and conduction bands are mostly of $p$ and $s$ character, respectively, yielding a zero winding number $\nu=0$. The definition of winding number and details of calculations can be found in Appendix \ref{windingsec}. 

As can be seen in Fig.~\ref{scheme}(b), the TI phase, interestingly, shrinks as the short-range interactions become stronger, until, eventually, the EI phase emerges at $U_{sp}\approx3$ for nearly zero on-site energies. Therefore, in strongly correlated systems, the EI phase surpasses the TI phase and we only have exciton and band insulator states. In an intermediate coupling strength, i.e., $2<U_{sp}<3$, all three topological phases can be reached by varying on-site potential $\epsilon$. In both TI and BI phases, where the exciton order parameter vanishes, $\phi=0$, the mean-field Bloch Hamiltonian has a chiral symmetry. This can be seen from Eq.~\eqref{b_spin} with the following Bloch Hamiltonian 
\begin{align}\label{bloch}
H^{\mathrm{MF}}_k=\mathbf{d}(k)\cdot\boldsymbol{\sigma},
\end{align}
where $\mathbf{d}(k)=1/2(B_k^{1},B_k^{2},B_k^{3})$, and also, $B_{k}^{1}=0$ when $\phi=0$. Hence, since $\{H^{\mathrm{MF}}_k, \sigma_1\}=0$, the system is manifestly chiral symmetric and the winding of unit vector $\hat{\mathbf{d}}(k)\equiv \mathbf{d}(k)/||\mathbf{d}(k)||$ around the origin in the $y-z$ plane determines $\nu$ as one crosses the one-dimensional Brillouin zone. Therefore, we obtain $\nu=-1$ for TI phase and $\nu=0$ for BI phase.  
 
%
\begin{figure}[t]
    \center
    \includegraphics[width=\linewidth]{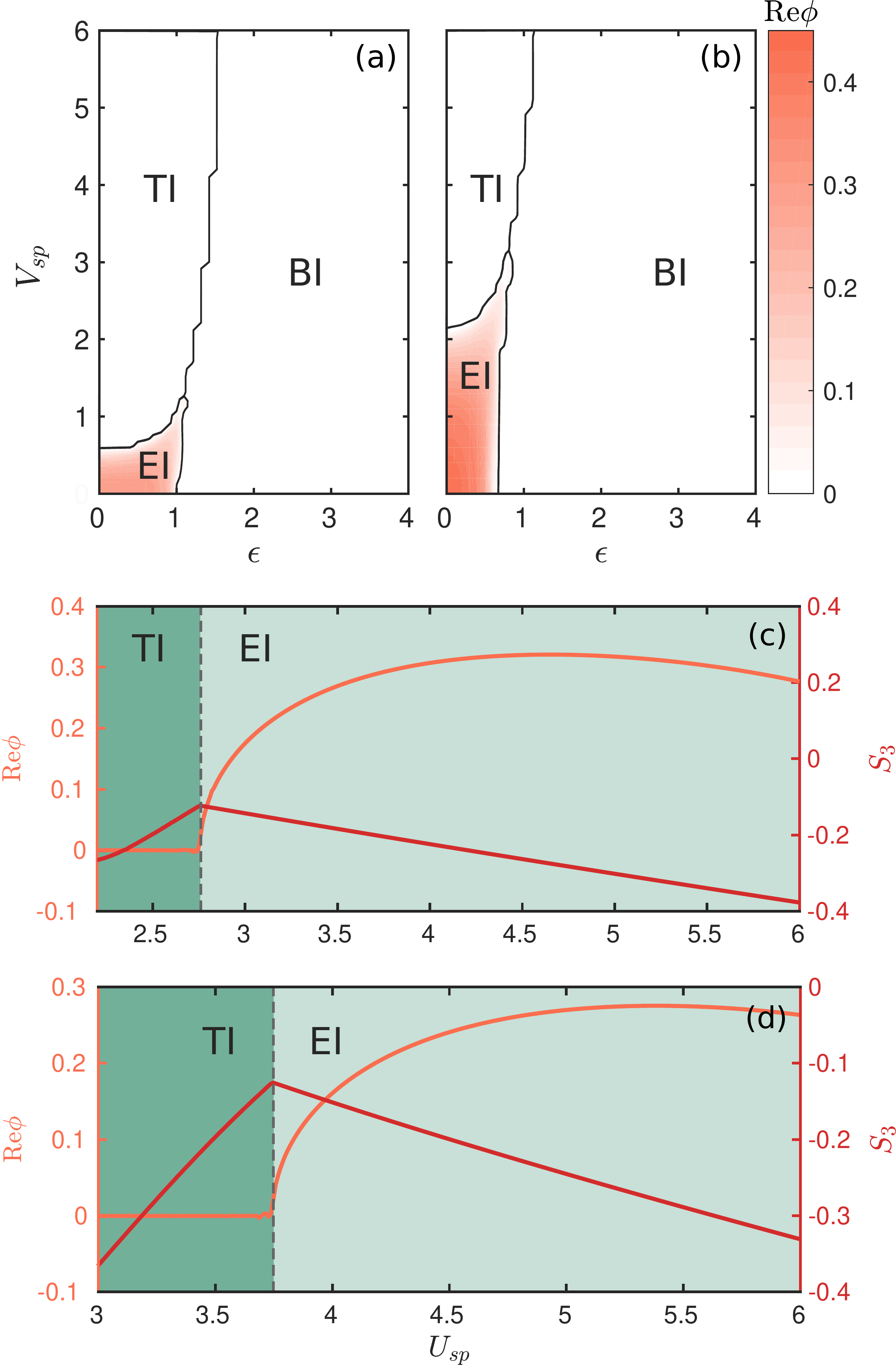}
    \caption{(a,b) Surface plots of equilibrium exciton order, as a function of on-site energy ($\epsilon$) and $s-p$ hybridization parameter ($V_{sp}$) for fixed Coulomb interaction (a) $U_{sp}=3$, and (b) $U_{sp}=6$. (c,d) Self-consistent solutions for exciton order (orange) and $z$ component of pseudospin (red) as a function of Coulomb interaction. The on-site energy is $\epsilon=1/2$ for both panels, whilst, the inter-orbital hopping parameter is set to (c) $V_{sp}=1/2$, and (d) $V_{sp}=1$, respectively. $S_3$ is proportional to $n_s-n_p$ and acts as an effective orbital pseudomagnetization. The susceptibility (i.e. the derivative of the pseudomagnetization) at a critical pseudomagnetic field (dashed line), becomes discontinuous, alluding a phase transition where excitons emerges. \label{scfphi}}
\end{figure}

In Fig.~\ref{scfphi}(a,b) the impact of the $s-p$ orbitals hybridization on the phase diagram is shown. The plots in panel (a) and panel (b) depict the topological behavior of $s-p$ chain for two specific values of $U_{sp}=3$ and $6$, respectively. As can be seen in both panels, the EI state is limited to the low hybridization region and enhances towards larger hybridization as the short-range interactions become stronger. Thus, larger the inter-orbital Coulomb interaction is, larger value of hybridization is required for the TI phase to set in.

The physical picture obtained thus far is that the EI phase emerges out of the TI phase in the strong interaction regime. Therefore, the two phases compete with each other. Moreover, we find that the EI phase is topologically {\it trivial}, and the nonzero exciton parameter leads to $B_{k}^{1}\neq0$ and, consequently, breaks the chiral symmetry. This implies that the unit vector $\hat{\mathbf{d}}(k)$ tips out of the $y-z$ plane, and thus the closed path traveled by $\hat{\mathbf{d}}(k)$ can be shrunken to zero continuously. From a mathematical viewpoint, in the EI phase, the latter unit vector belongs to the surface of a sphere $S^2$, and since the first homotopy group of $S^2$ is trivial $\pi_1(S^2)=0$ \cite{Nakahara2003}, the phase is trivial.

\subsection{Phase transition to EI phase}
In order to understand the phase transition to EI, we present an analytical study of order parameters. To this end, we build the mean-field Hamiltonian based on pseudospin components, followed by calculation of the exciton parameter, so that we find the criteria at which the EI phase could emerge. From Eq.~\eqref{phi-n0-n1}, we obtain the exciton parameter at the half filling state, i.e. $ f (E_{ k,+}(0)) - f (E_{k,-}(0))=-1$, as
\begin{align}\label{phisc}\nonumber
\phi&=\frac{1}{N} \sum_k\left(\langle S^1_k\rangle+i\langle S^2_k\rangle  \right)\\
&=\frac{1}{N} \sum_k\left(U_{sp}\phi+2iV_{sp}\sin k   \right)/\left| {\bf B}_k \right|.
\end{align}

To further simplify the above equation, we expand the norm of the pseudomagnetic field vector,
\begin{align}\label{normb}
    \left| {\bf B}_k\right|=\sqrt{4U_{sp}^2|\phi|^2 +16V_{sp}^2\sin^2k+\left(2\epsilon-4 t\cos k  -2U_{sp}~S_3\right)^2}
\end{align}
which in fact, involves the evaluation of $S_3=1/N\sum_k\langle S^3_k\rangle$,

\begin{align}\label{S3}
S_3 =\frac{1}{2\pi}\int_{-\pi}^{\pi}dk
\frac{-2\epsilon+4 t  \cos k +2U_{sp}S_3}{\left| {\bf B}_k\right|},
\end{align}
that practically plays the role of a pseudomagnetic order parameter. Since $\left| {\bf B}_k\right|$ is an even function of $k$, Eq.~\eqref{phisc} simplifies to 
\begin{align}\label{eq1}
\phi=\frac{1}{N}\sum_k
\frac{U_{sp}\phi}{\left| {\bf B}_k\right|},
\end{align}
which is not dissimilar to the Bardeen-Cooper-Schrieffer (BCS) superconductor gap equation. Eventually, we have two equations for $\phi$ and $S_3$ that could be solved self-consistently to obtain a concrete condition over which the exciton parameter becomes nonzero. The results are displayed in Fig.~\ref{scfphi}(c,d) for the on-site energy $\epsilon=1/2$, and inter-orbital hopping parameter $V_{sp}=1/2$ (panel (c)), and $V_{sp}=1$ (panel (d)). As can be seen, a slope discontinuity occurs in $z$ component of pseudospin, $S_3$, which presents the difference between $s$ and $p$ orbitals charge density ($n_s-n_p$), acting as an effective pseudomagnetization in the orbital basis. From Eq.~\eqref{phisc} and \eqref{S3}, it is clear that $\phi$ and $S_3$ make the in-plane and out of plane components of the pseudomagnetization. With this identification, the inter-orbital Coulomb interaction $U_{sp}$ acts like a magnetic field. Hence, it's seen that $\partial S_{3}/\partial U_{sp}$ becomes discontinuous at a critical value of interaction alluding a phase transition to an ordered phase, i.e EI. The nonzero $\phi$ amounts to developing an $x$-component of pseudomagnetization. This is exactly the tipping of $\hat{\mathbf{d}}(k)$ out of the $y-z$ plane as we mentioned in the preceding subsection. Note that the creation of $x$-component is set by hybridization of $s-p$ orbitals which is facilitated in the band-inverted TI phase. Thus, contrary to previous studies, in a BI phase in our model, no band mixing occurs which is detrimental in forming nonzero $\phi$ even for strong interactions.      

\subsection{BCS-BEC crossover in the EI phase}
As we showed above, the strong inter-orbital Coulomb interaction can establish the EI phase in the $s-p$ model. Our results further reveal that there is a BCS-BEC crossover within the EI phase by varying interaction and hybridization strength. To identify the crossover we use the shape of the bands as an estimation of BCS and BEC phases. For a more pricise identification of phases one has to use the size of excitons \cite{tanabe2018nonequilibrium, Kaneko2012, kunevs2015excitonic}. In the BCS regime, the momentum of maxima of valence band, $k_F$, appears away from $k=0$, reflecting the existence of weakly bound excitons at finite $k$. When moving toward stronger interaction regime, the Hartree potential broadens the valence and conduction band densities and pushes the maxima to $k=0$. For bands being flattend around the center of BZ, we use the increasing behavior of $\phi$ by $U_{sp}$ from BCS, bands with two minima and maxima, to charactrize the crossover to the BEC phase. When $\phi$ start to decrease, the BEC phase sets in (see Fig.~\ref{spbands2} in Appendix \ref{bcs-bec}). 

In Fig.~\ref{fig3} we present a phase diagram in $U_{sp}$-$V_{sp}$ plane where the domain of BCS and BEC phases is indicated. BCS phase can be only found in a low $s-p$ hybridization energy limit and by increasing the short-range interaction strength, the bands are flattened until a BCS-BEC crossover occurs. In Appendix \ref{bcs-bec} we present the details of band evolution, which clearly shows how the crossover takes place. 

The BCS-BEC crossover has been addressed very recently in a three-orbital model with the inclusion of spin-orbit coupling (SOC) \cite{kaushal2020bcs}. It's been shown that the SOC-originated condensation of BCS type EI at intermediate Coulomb interaction region crosses over to a BEC type EI by increasing SOC strength. This suggests that SOC in their model acts analogously to the hybridization of the $s$ and $p$ orbital in our model. Furthermore, in a very recent study, the ${\rm Ta_2NiSe_5}$ phase transition that was observed as an anomaly in the resistivity at $T_c\simeq328$K \cite{Salvo1986physical}, has been modeled via 1D EFKM where a BEC type condensation occurs even though the normal state is an overlapping band semimetal \cite{Sugimoto2018strong}. This in fact alludes to the notion of BCS-BEC crossover.

\begin{figure}[t]
	\center
	\includegraphics[width=\linewidth]{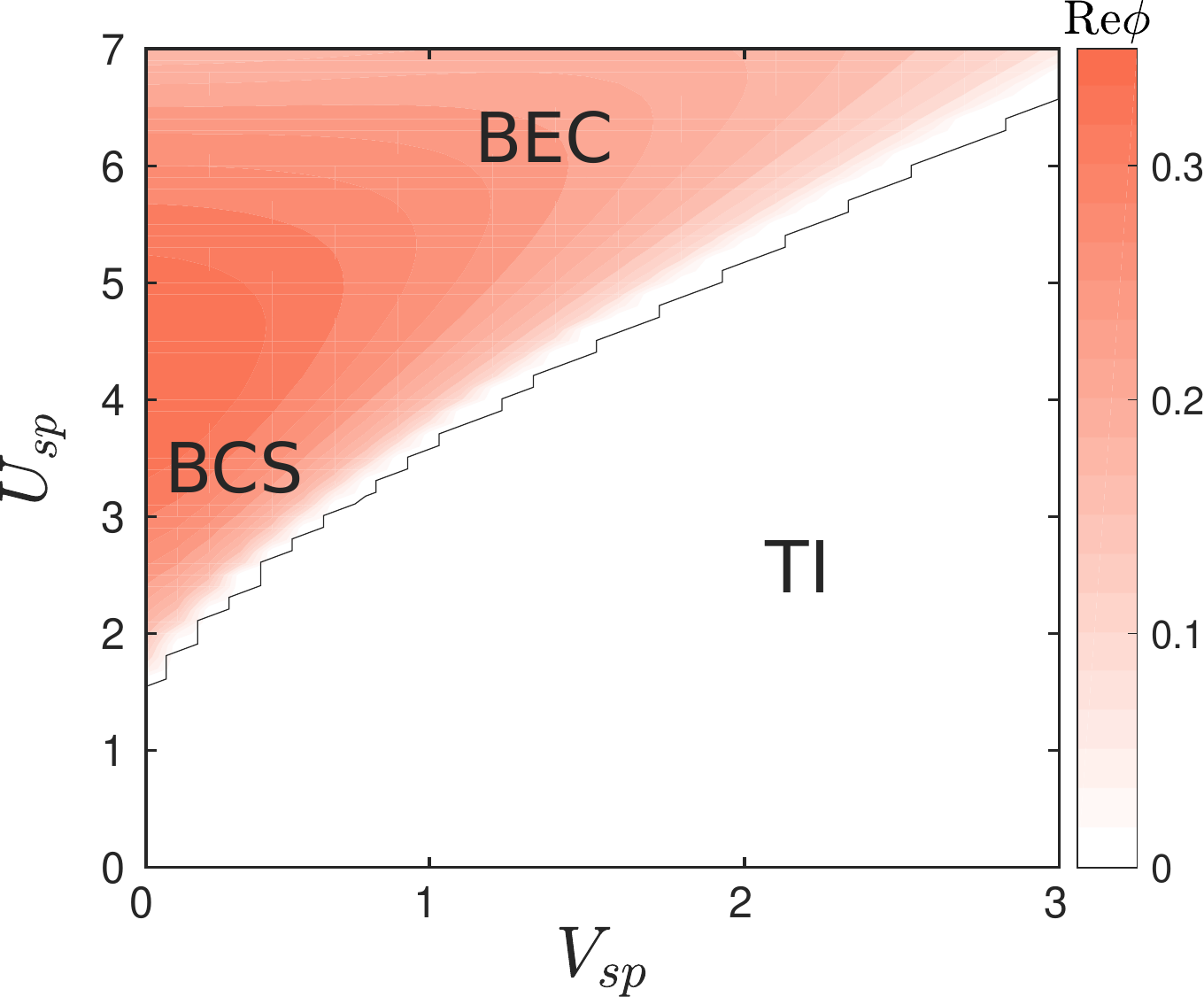}
	\caption{The exciton order of a 1D $s-p$ chain at equilibrium with on-site energy of $\epsilon=1/2$. The BCS-BEC crossover is visible at low values of $V_{sp}$. The details of BCS-BEC crossover with Coulomb coupling enhancement is shown and discussed in Fig.~\ref{spbands2}.\label{fig3}}
\end{figure}

\subsection{Collective modes}\label{CMs}

\begin{figure}[t]
	\center
	\includegraphics[width=\linewidth]{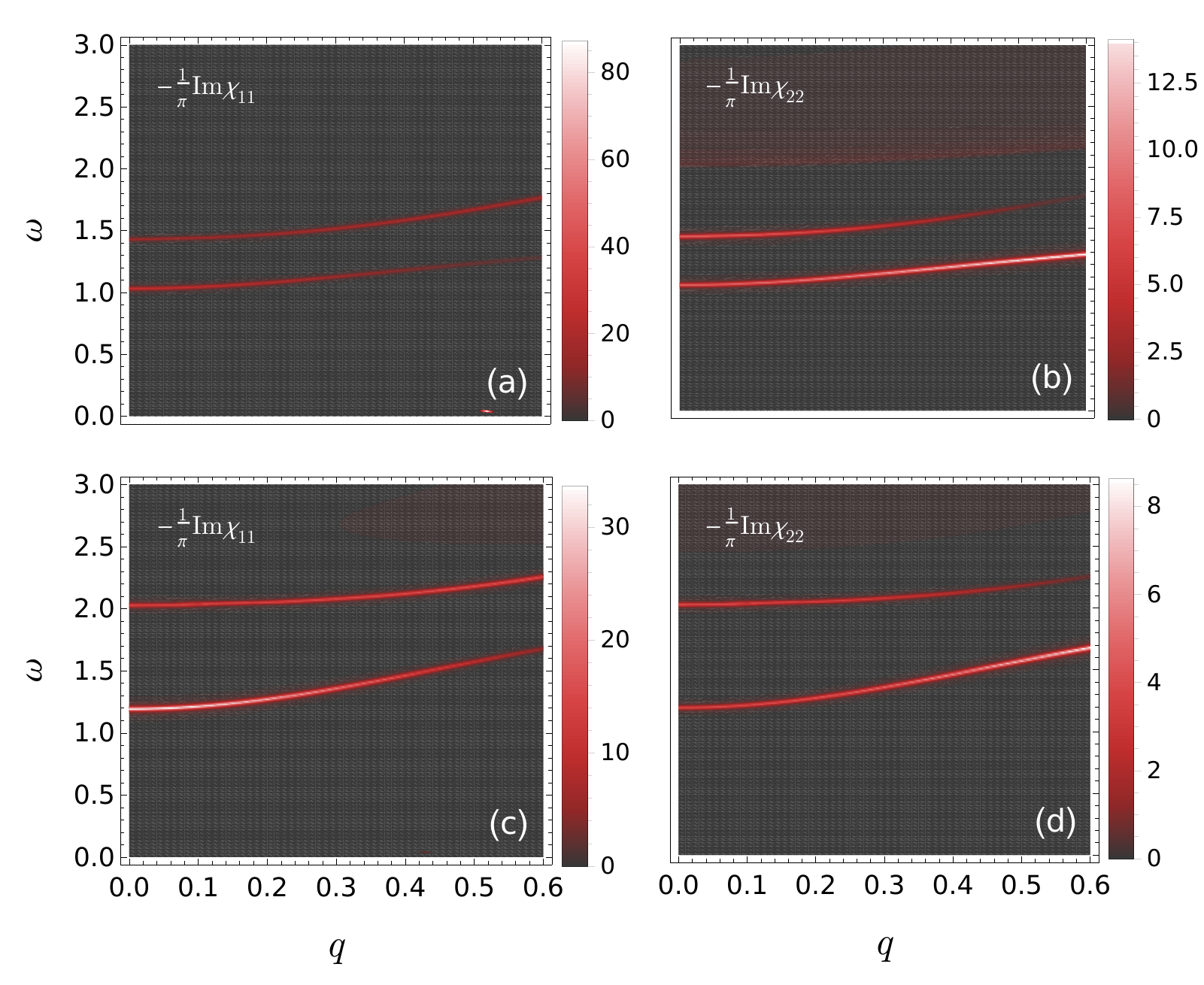}
	\caption{(a,c) Amplitude mode and (b,d) phase mode of the BCS and BEC exciton insulator. Results for a BCS EI with $\epsilon=1/2$, $V_{sp}=1/2$, $U_{sp}=2.8$, are illustrated in (a,b) and those for a BEC EI with $\epsilon=1/2$, $V_{sp}=1$, $U_{sp}=3.8$ are presented in panel (c) and (d). In all panels we set $g\approx0.07$ for electron-phonon coupling strength. \label{suscep}}
\end{figure}

At equilibrium the mean-field ground state of the EI phase is characterized by the order parameter $\phi$ discussed in preceding sections. The fluctuations of amplitude and phase of the order parameter create collective modes. To study the latter modes, we condenser the retarded density correlation function for the $s-p$ chain using the Kubo formula,

\begin{align}\label{susceptibility}
\chi^R_{\mu\nu}(t) =-i\theta(t) \frac{\langle\Psi_0|[\rho_\mu(t),\rho_\nu(0)]|\Psi_0\rangle}{\langle\Psi_0|\Psi_0\rangle}.
\end{align}

Here, $\rho_\mu\equiv \frac{1}{N}\sum_{k}\zeta_k^\dagger \sigma_\mu \zeta_k$ is the collective charge and exciton modes, and $|\Psi_0\rangle$ is the ground state of the interacting system. The perturbative expansion of Eq. \eqref{susceptibility} can be evaluated with the help of Wick's theorem in the interaction picture. By Fourier transformation to momentum and frequency domain, and within the  random-phase approximation (RPA),  Eq. \eqref{susceptibility} can be cast as

\begin{align}
\chi^R(\omega,{\bf q}) =\chi^0(\omega,{\bf q})+\chi^0(\omega,{\bf q}){\cal U}~\chi^R(\omega,{\bf q}),
\end{align} 
where $\chi^0(\omega,{\bf q})$ is the bare susceptibility and ${\cal U}=(U_{sp}/2){\rm diag}(1,-1,-1,-1)+ {\rm diag}(0,D,0,0)$. Here, $D=g^2D_0/(1-g^2\chi^0_{11}D_0)$ and $D_0=2\omega_{ph}/((\omega+i0^{+})^2-\omega_{ph}^2)$ are the dressed and bare phonon propagators, with $g$ as the electron-phonon coupling and $\omega_{ph}$ being the frequency of optical phonons. The details of phonon Hamiltonian and electron-phonon coupling are discussed in next section. 

The results for RPA susceptibility is shown in Fig \ref{suscep} for both BCS and BEC exciton condensates. Here, we have set $\omega_{ph}=0.1$ and used the broadening factor $\eta=0.01$.
Each row indicates the collective modes in the amplitude ($-\chi^R_{11}/\pi$) and phase ($-\chi^R_{22}/\pi$) direction for a specific $s-p$ chain. Fig. \ref{suscep}(a) and Fig. \ref{suscep}(c) clearly indicate the gapped nature of the amplitude mode, the lower excitation branch starting from a finite value at $q=0$. The upper branch is the onset of continuum of excitaions across the band gap.   

The results for the dispersion of phase modes are shown in Fig. \ref{suscep}(b) and Fig. \ref{suscep}(d) for BSC and BEC condensates, respectively. It is clearly seen that the lower dispersive branch is also gapped. Indeed, in the model used in our paper the phase and amplitude modes are coupled to each other due to the inter-orbital coupling $V_{sp}$, and consequently the gapped amplitude modes result in massive phase modes in contrast to collective modes of a EFKM EI \cite{murakami2017photoinduced}.

%

\begin{figure*}[!htb]
    \center
    \includegraphics[width=0.88\linewidth]{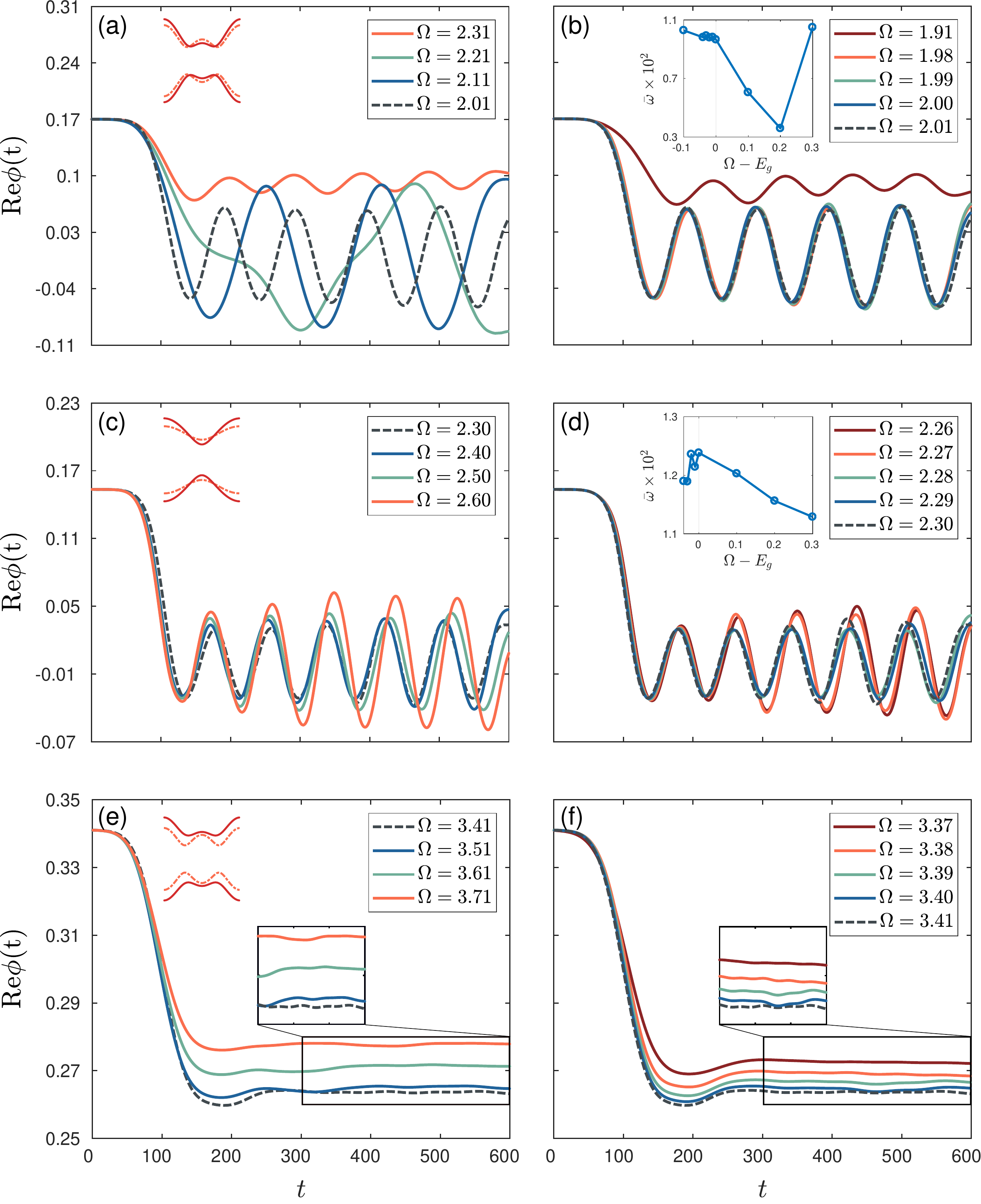}
    \caption{Time evolution of the exciton parameter for diverse $s-p$ chain systems. (a,b) illustrates the results for a $s-p$ chain with $\epsilon=1/2, V_{sp}=1/2, U_{sp}=2.8$, while panel (c,d) shows those for $\epsilon=1/2, V_{sp}=1, U_{sp}=3.8$. (e,f) are exciton dynamics for $\epsilon=1/4, V_{sp}=1/2, U_{sp}=4$. Black dashed line indicates the exciton order time evolution for laser frequency $\Omega=E_g$. For both EI phases, (a,b,e,f) BCS and (c,d) BEC, the exciton order melts down during the exposure of laser pump and oscillates afterward. In panel (e,f) the exciton dynamics in $t\in [300,600]$ and $\mathrm{Re}\phi(t)\in[0.26,0.28]$ window are magnified to ease the visualization. The left (right) column insets are the corresponding band dispersion (oscillation frequency). The dashed (solid) lines depict energy dispersion for non-interacting $U_{sp}=0$ (interacting  $U_{sp}\ne 0$) $s-p$ chain in the inset. Phonon-charge interactions addressed in \ref{phonon_coupling}, are considered in preparation of this figure with the following assumptions. $g=\sqrt{\lambda\omega_{ph}/2}$ is the charge-phonon interaction factor. $\lambda=0.1$ is the effective electron-phonon coupling parameter, and $\omega_{ph}=0.1$ is the optical phonon energy. Note that since the imaginary part of the condensate is negligibly small, here we only present $\Re\phi$.
    	\label{dynamsp}}
\end{figure*}

\section{Excitons in non-equilibrium state}\label{noneq}
Motivated by recent pump-probe measurements on flakes of ${\rm Ta_2NiSe_5}$ \cite{andrich2020imaging}, in this section we study the non-equilibrium dynamics of the EI phase of the $s-p$ model described in the preceding section. The non-equilibrium dynamics of excitons in a 1D $s-p$ chain generated with a laser pump provides an excellent playground for understanding the collective behaviors. We first present the details of the studied non-equilibrium model, then we elaborate on the real-time evolution of exciton condensate. Note that, intra-atomic hybridization between non-localized $s$ and $p$ orbitals leads to the dipolar transitions \cite{Pedersen2001}, i.e. $H_{dip}=E(t) \sum_j (c^{\dagger}_j p_{j}+ c^{\dagger}_{j} p_{j})$, and consequently exciton order renormalization due to modification of pseudomagnetic field component,

\begin{align*}
B^1_k=-2 U_{sp} \Ree{\phi}+2gX+2E(t),
\end{align*}
where, $E(t)=-\frac{\partial }{\partial t}A(t)$ is the laser pulse electric field, and $A(t)$ is its electromagnetic vector potential. However, for the sake of simplicity we neglect the dipolar transitions.

\subsection{Coupling to phonons and laser pulses}\label{phonon_coupling}
We consider the charge interactions with a bath of optical phonons with $\hbar\omega_{ph}$ energy. The Hamiltonian is modified as $H=H_{e}+H_{ph}$, where 
\begin{align}
H_{ph}=\hbar\omega_{ph} \sum_{j} b_{j}^\dagger b_{j}+g \sum_{j} (b_{j}^\dagger + b_{j} )(c^\dagger_j p_j+p^\dagger_j c_j),
\end{align}
$g$ is the electron-phonon coupling, and $b^{\dagger}_j (b_j)$ is the creation (annihilation) operator for phonons. The charge-phonon interaction in mean-field approximation reduces to
\begin{align}
H_{ph}^{\rm MF}= &~\hbar\omega_{ph} \sum_{j} b_{j}^\dagger b_{j}+g X \sum_j ( c_{j}^\dagger p_{j}+p_{j}^\dagger c_{j})\nonumber\\
&+ g \sum_j ( b_{j}^\dagger+ b_{j})(\phi+\phi^\ast),
\end{align}
where $X=\langle b_{j}^\dagger+ b_{j}\rangle$ is the phonon displacement.
The total Hamiltonian in momentum space thus can be written as the summation of Fourier transform of electron-phonon mean-field Hamiltonian and Eq.~\eqref{b_spin},
\begin{align}\label{Hmf}
H^{\rm MF}=\bar{H}_{e}^{\rm MF}+\hbar\omega_{ph} \sum_{j} b_{j}^\dagger b_{j}+2 g \Re{\phi}\sum_j ( b_j^\dagger+ b_j)
\end{align}
with $\bar{H}^{\rm MF}_e$ being the electronic part of the mean-field Hamiltonian with a slightly modified pseudomagnetic field component $B^1_k=-2 U_{sp} \Re{\phi}+2gX$. 

Next, we model an optical laser pump pulse impinging on the system. We assume the induced time-dependent electromagnetic vector potential as a Gaussian function

\begin{equation}\label{A-pulse}
{\bf A}(t)=\Theta(t)~A_0~e^{-\frac{(t-t_p)^2}{2\tau_p^2}}\sin(\Omega t/\hbar).
\end{equation}

 Here, we set the pulse amplitude ($A_0$) to $0.05$ and $\hbar=1$ throughout this paper. We also set $t_p=100$ and $\tau_p=30$ as the duration and width of the pulse, respectively. The non-equilibrium state could be modeled by a Peierls phase in the mean-field Hamiltonian \cite{tanabe2018nonequilibrium}. The Heisenberg equation of motion provides the time evolution of electron-hole pairs, 

\begin{subequations}\begin{align}\label{hme_x}
    \frac{\partial \left\langle {\bf S}_k (t) \right\rangle }{\partial t}
    &=\bm{B}_k (t) \times \left\langle \bm{S}_k (t) \right\rangle
    \\
    \frac{\partial \left\langle S^0_k (t) \right\rangle }{\partial t}&= 0\\
    \frac{\partial X (t) }{\partial t}&=\omega_{ph}P(t) \label{ph-dis}\\
    \frac{\partial P(t) }{\partial t}&=-\omega_{ph} X(t) -4 g\Re{\phi}.\label{pt}
    \end{align}\end{subequations}

Here, $P(t)=i\langle b_j^\dagger-b_j\rangle$ is phonon momentum.
We solve the set of above equations, \eqref{hme_x}-\eqref{pt}, numerically by Runge-Kutta fourth-order method where we insert the self-consistent results as the initial value for exciton order parameter. We assume that the system is at zero temperature and thus the initial phonon momentum vanishes $P(0)=0$. From Eq.~\eqref{pt}, this assumption yields $X(0)=-4g\phi(0)/\omega_{ph}$. At equilibrium ($t\leq0$) the pseudomagnetic field component discussed below Eq.~\eqref{Hmf}, becomes $B_{k}^{1}=-2(U_{sp}+2\lambda)\mathrm{Re}\phi$ with $\lambda\equiv 2g^2/\omega_{ph}$ being the electron-phonon coupling constant. Hence, at equilibrium the latter interaction only shifts $U_{sp}$ \cite{Murakami2020Collective}. In the following we discuss the evolution of system for $t>0$ irradiated by the pump pulse \eqref{A-pulse}.   

\subsection{Real-time evolution of EI condensate}

The time evolution of the exciton order is illustrated in Fig.~\ref{dynamsp}. Panels (a) and (b) show a BCS type EI dynamics, whilst panel (c) and (d) depict those for a BEC EI. All EI phases demonstrate an exciton order melt-down after the laser pump exposure as a consequence of photo-induced breaking of exciton bound states. In each panel, we show the real-time evolution of Re$\phi$ for different values of pulse frequency $\Omega$. The black dashed line shows Re$\phi(t)$ when $\Omega$ equates the EI gap energy. From panel (b) we see that, when the pulse frequency is lower than the gap, the oscillation of condensate is almost the same for all. However, for $\Omega$ being larger than the gap, the oscillations do depend on pulse frequency. This behavior can be ascribed to the type of band structure associated with the BCS condensate illustrated in the inset of panel (a). The variation of band structure near $k=0$ introduces different resonance energy scales for the condensate. For frequencies well above the gap, these variations are smeared out; the same reasoning holds for frequencies below the gap as shown in panel (b).

The oscillation frequency $\bar{\omega}$ of Re$\phi(t)$ with respect to pulse frequency is depicted in the inset of panel (b). 
Note that in all latter cases, $\bar{\omega}$ of exciton condensate coincides with the phonon frequency, i.e., with the oscillation of $X(t)$. We note that in the absence of electron-phonon coupling the atoms oscillate with their natural optical frequency which we set to be $\omega_{ph}=0.1$ throughout. However, when coupled to exciton condensate the phonon frequencies change. The coupling between phonons and excitons has been argued to be crucial in understanding the recent pump-probe measurements on ${\rm Ta_2NiSe_5}$ \cite{Werdehausen2018,andrich2020imaging,Murakami2020Collective}.           

The results of the condensate dynamics for BEC type EI are shown in Fig.~\ref{dynamsp}(c,d). The band structure depicted in panel (c) clearly shows that the direct gap is located at $k=0$. Again in this case the exciton order parameter is quenched by the pump pulse and oscillates afterward. The main observation now is that for pulse frequencies either below or above the bandgap, the exciton frequencies $\bar{\omega}$ changes only mildly in contrast to the BCS case discussed above. The reason can be ascribed to the band structure of the BEC with only one direct bandgap at $k=0$. The non-equilibrium dynamics of BEC condensate in our model is consistent with that of the EFKM BEC, meaning that the exciton condensate undergoes a photo-induced suppression. However, when moving energetically far away form the bandgap, we do not observe any exciton condensate enhancement \cite{tanabe2018nonequilibrium}. The suppression of condensate by the photoexcitation for a BCS type EI is reported in a two-orbital Hubbard model, while the photo-induced enhancement of condensate is predicted for BEC condensate\cite{Tanaka2018}. 

Finally, in Fig.~\ref{dynamsp}(e,f), we present the results for non-equilibrium dynamics of condensate in an EI with large bandgap $\sim 3.41$. Two observations are manifest. First, we see that the condensate is less influenced by the laser pulse than the other aforementioned cases. That is, Re$\phi$ drops to a small fraction of its initial value; it changes from Re$\phi\simeq0.34$ to Re$\phi\simeq0.26$ even at the resonance with the gap energy. Second, in stark contrast to previous cases, no oscillation occurs for a full range of pulse frequencies below and above the EI gap. Compared to dynamics in panels (a-d), one can see that the effective coupling of condensate to phonons only takes place in the EI phase with a narrow gap.

Before moving to the next section, a remark is in order. The condensate dynamics saturates to a mean steady value after a melt-down and does not decay, which can be attributed to the lack of damping mechanism induced by phonons, contamination, doping, and defects in general. This is actually the motivation behind the recent time-resolved photoluminescence measurements reporting a high exciton lifetime in encapsulated TMDs \cite{fang2019control}. In our model the decay could be considered by adding a damping term as $-\gamma P(t)$, with $\gamma$ as a damping parameter, phenomenologically \cite{murakami2017photoinduced} to the right side of Eq.~\eqref{pt}, which leads to decay of oscillations.


%
%
%
\begin{figure}[t]
    \center
    \includegraphics[width=\linewidth]{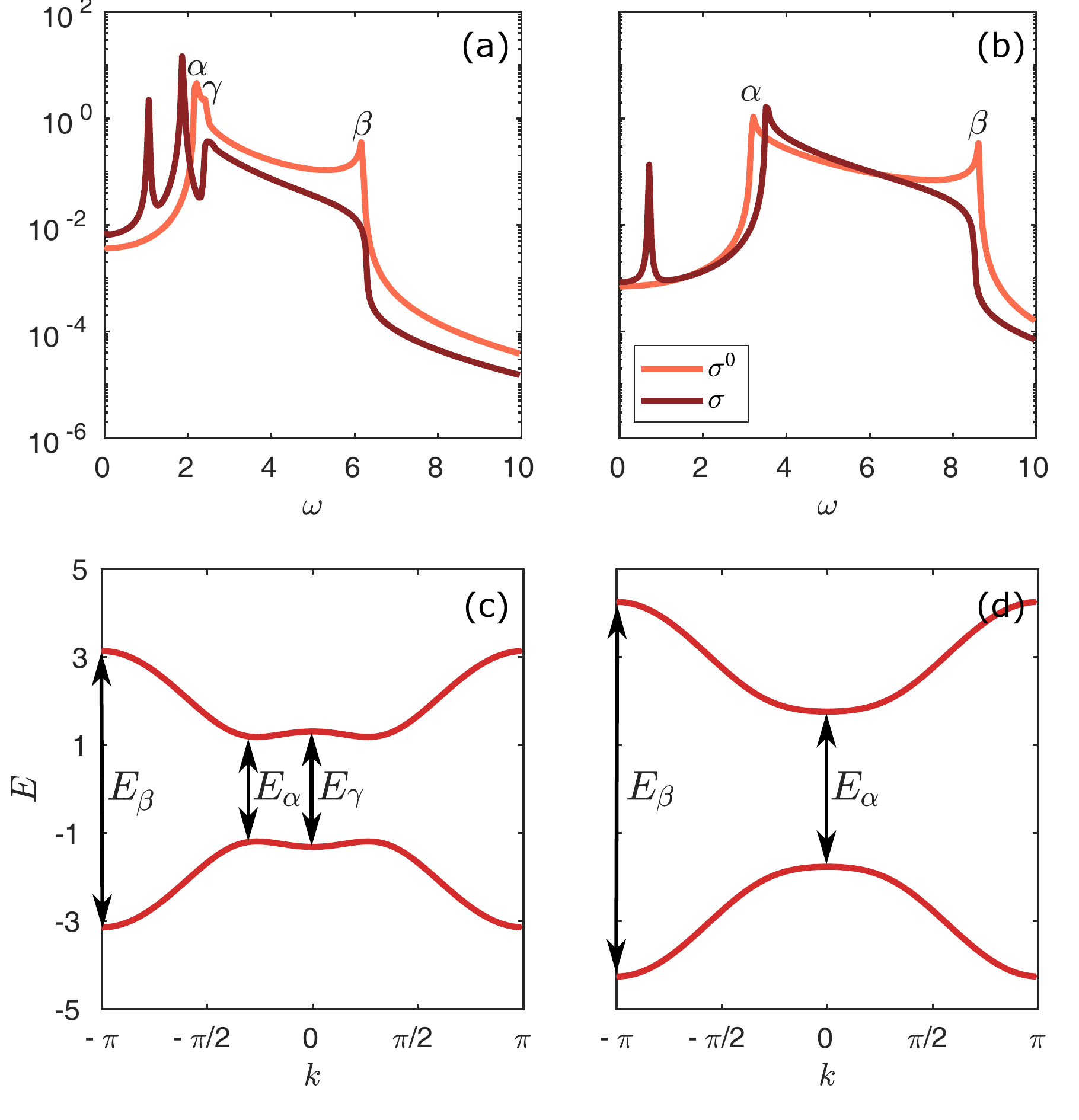}
    \caption{
        (a, b) Optical conductivity and (c, d) mean-field energy dispersion of 1D $s-p$ chain at $t=0$ (equilibrium state). Panel (a, c) depict results for $\epsilon=1/2$, $V_{sp}=1/2$, $U_{sp}=3.2$, while (b, d) illustrates those for $\epsilon=1/2$, $V_{sp}=1/2$, $U_{sp}=5$. In (a, b) the orange curve indicates the real part of the bare optical conductivity $\sigma^{0}(\omega)$, and the optical conductivity including vertex corrections, $\sigma(\omega)$, is shown in dark red. In (c) and (d) black arrows depict the optical transitions with strongest contribution to the bare optical response. The appearance of many-body optical transition associated with the collective modes is manifest as a sharp peak in the single-particle gap. We set $g\approx0.07$ for electron-phonon coupling.
        \label{opcosp}}
\end{figure}
%
%
%

%
%
\begin{figure*}[!htb]
    \center
    \includegraphics[width=\linewidth]{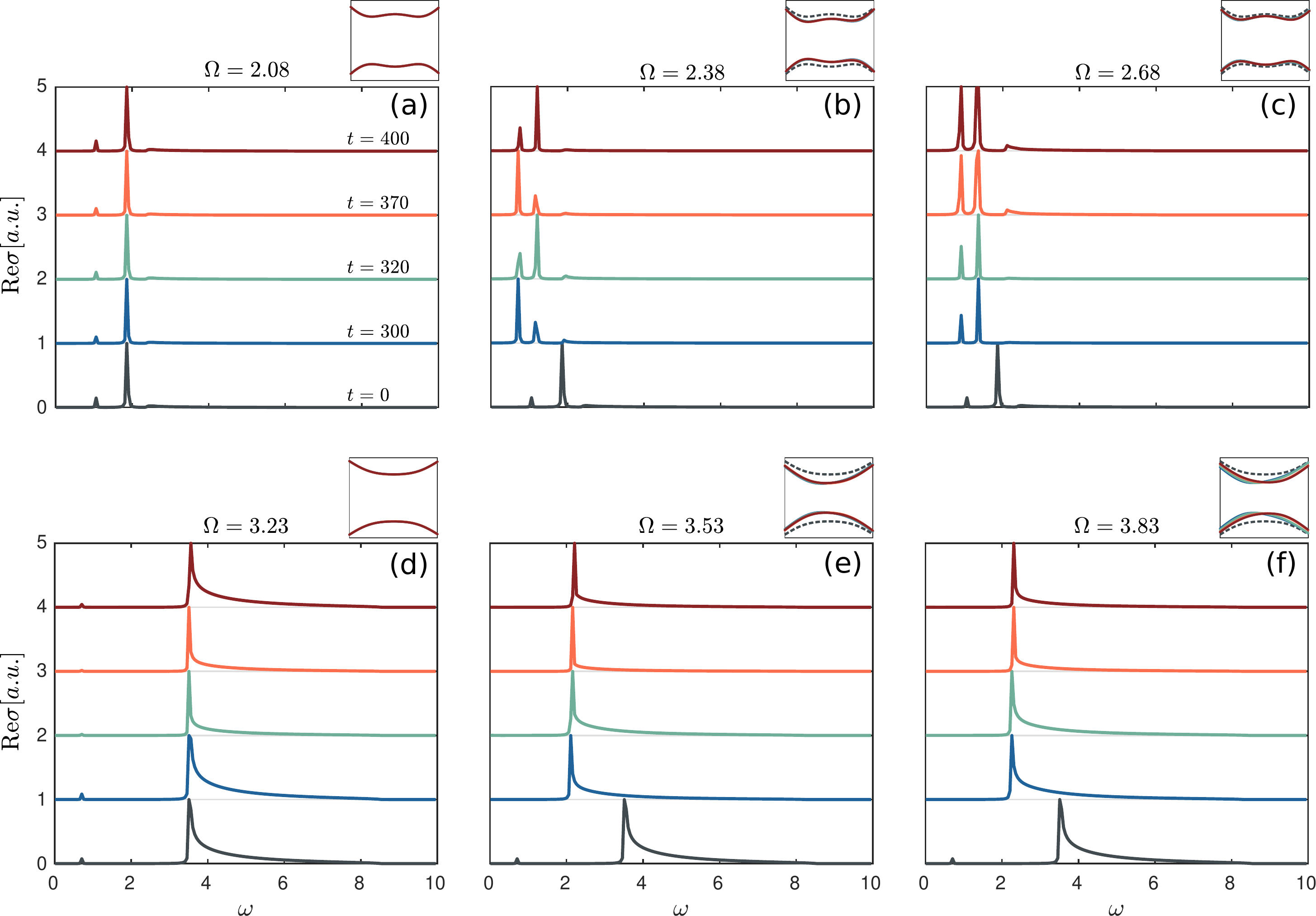}
    \caption{
        Optical response of (a-c) BCS, (d-f) BEC EI systems introduced in Fig.~\ref{opcosp}, for finite time after the laser pulse exposure. Each panel presents the optical conductivity for specific laser pump frequency ($\Omega$). Black bottom line in each panel is the optical spectra for the corresponding EI at equilibrium. The energy dispersion at times shown in panels, are plotted for comparison (See top right of each panel). The dashed line indicates the energy dispersion at equilibrium. The electron-phonon coupling ($\lambda$) is set to $0.1$ in all panels. 
        \label{nonopcosp}}
\end{figure*}

\section{Optical conductivity}\label{opcosec}
This section aims to answer the last question posed in introduction seeking the signature of EI phases in optical measurements. The longitudinal optical conductivity can be found using the relation $\sigma(\omega)=\lim\limits_{q\to0}\frac{i}{\omega}\Pi^R(\omega,{\bf q})$, where $\Pi^R(\omega,{\bf q})$ is the retarded current-current correlation function. The optical conductivity can be written as $\sigma(\omega)=\sigma^{0}(\omega)+\sigma^{v}(\omega)$ where $\sigma^{0}(\omega)$ is the bare optical conductivity and $\sigma^{v}(\omega)$ includes the vertex corrections due to inter-orbital Coulomb interaction and electron-phonon coupling. The bare part reads as 

\begin{align}
\sigma^0(\omega)=\frac{i}{\omega}\int\frac{dk}{2\pi}\left(\frac{1}{\omega-\omega_{k}+i0^+}-\frac{1}{\omega+\omega_{k}+i0^+}
\right)|{\cal J}|^2,
\end{align}
where $\omega_{k}=E_{k,+}-E_{k,-}$, $\omega$ is the probe frequency, and ${\cal J}$ is the current matrix element between the conduction and valence bands, 
{\small
\begin{align}\label{currentden}
{\cal J}=-2e\left(t\sin k \frac{B^1_k+iB^2_k}{|{\bf B}_k|}-i\frac{V_{sp}}{2}\cos k\left\{ \left(1+\frac{B^3_k}{|{\bf B}_k|}\right)e^{2i\theta}+\left(1-\frac{B^3_k}{|{\bf B}_k|}\right)\right\}\right)
\end{align}}
with $\theta=\arctan B^2_k/B^1_k$. From Eq.~\eqref{currentden} we see that, the current matrix element is a function of exciton parameter ($B^1_k+iB^2_k$) and the $s-p$ orbitals hybridization parameter ($V_{sp}$). Also, via setting $V_{sp}=0$, one can reproduce the relation for current density operator of a 1D EFKM \cite{tanabe2018nonequilibrium}, ${\cal J}=-2et\sin k (B^1_k+iB^2_k)/|{\bf B}_k|$. We also note that when both $\phi$ and $V_{sp}$ are zero, optical conductivity vanishes, since ${\cal J}=0$. However, when the exciton condensation is formed, the matrix element becomes nonzero and optical response acquire finite values as a function of measured frequencies.

The real part of the bare optical conductivity $\sigma^{0}(\omega)$ is shown in Fig.~\ref{opcosp} for diverse values of Hamiltonian parameters at equilibrium. Fig.~\ref{opcosp} (a) and (b) present the longitudinal optical spectra for BCS and BEC EIs, respectively. Note that both plots demonstrate condensations in the TI phase. From the optical spectra, one can clearly see that in the linear response regime denoted in orange color, more than one excitonic bound state form, including at exact bandgap energy transitions ($E_\alpha$), and the BZ edge transitions ($E_\beta$ in (c) and (d)). Thus, the exciton order, at equilibrium state, is a superposition of all bound states. There also exists an additional peak in the optical absorption of BCS type EI which stems mostly from the zero momentum transitions ($E_\gamma$). 

The vertex corrected part of conductivity, $\sigma^{v}(\omega)$, is  

\begin{align} 
\sigma^{v}(\omega)=\sum_{\mu\nu}\int \frac{dk}{2\pi}\frac{dk'}{2\pi} \Gamma_{\mu}(k', \omega)V_{\mu\nu}(\omega)\Gamma_{\nu}(k, \omega),
\end{align}
with the vertex $ \Gamma_{\mu}(k, \omega)$ reading as
\begin{align}
\Gamma_{\mu}(k, \omega)=\sum_{\alpha\beta\gamma}\frac{\partial B^{\gamma}_k}{\partial k} \frac{f(E_{k,\alpha})- f(E_{k,\beta})}{\omega + E_{k,\alpha} -E_{k,\beta} + i0^{+}}\sigma^{\gamma}_{\alpha\beta}\sigma^{\mu}_{\beta\alpha},
\end{align}
where $\sigma^{\mu}_{\alpha\beta}=\langle \alpha|\sigma^{\mu}|\beta \rangle$ are matrix elements of Pauli matrices w.r.t eigenstates of mean-field Hamiltonian \eqref{b_spin}. The effective interaction with polarization insertion is as follows:
\begin{align}
V(\omega)=\left[1 - \mathcal{U} \chi^{0}(\omega) \right]^{-1}\mathcal{U}.
\end{align}

The vertex corrections change the optical transitions substantially. The optical conductivity $\sigma(\omega)$ is shown by dark red color in Fig.~\ref{opcosp} (a)-(b). It is clearly seen that the interactions smear out the single-particle $\beta$ peak for both BSC and BEC phases. The single-particle transitions at $\alpha$ and $\gamma$ are also shifted. Besides, an additional many-body peak appear at low frequencies which is associated with the collective modes described in Sec. \ref{CMs}.

We now utilize a similar procedure to evaluate the optical absorption spectra in a photo-induced excited regime (non-equilibrium state after the imposition of a laser pump). Fig.~\ref{nonopcosp} represents the real part of longitudinal optical conductivity for (a-c) BCS and (d-f) BEC EI previously discussed in Fig.~\ref{opcosp}. Each column depicts the optical response to a specific laser pump energy ($\Omega$).
The bottom plot marked in black in each panel depicts the equilibrium optical response for the corresponding system. As previously discussed, single-particle peaks and also many-body transition appear in the equilibrium optical response. The single-particle peaks stem from the optical transitions close to the BZ center (black bottom curves in panel (a-c)). 

At a finite time far away from the laser pump exposure time ($t_p$), when the pulse energy is lower that the bandgap, we see that all excitonic transitions remain almost intact indicating that the energy bands are not significantly altered in both BCS and BEC EIs as shown in Fig.~\ref{nonopcosp} (a) and (d). 

On the contrary, for the laser pulses with equal and larger energies compared to bandgap, in the BEC EI shown in Fig.~\ref{nonopcosp} {\color{red}(e,f)}, we observe that the manybody excitations are quenched while the single-particle excitations redshift to lower energies as a consequence of photo-induced bandgap shrinkage. This is clearly seen in the band dispersion shown in the top right of each panel; the energy dispersion (solid lines) at finite times compared to one at equilibrium (dashed line). Note the time scale over which the exciton order parameter changes, which is about $\bar{\omega}^{-1}\sim 100~\mathrm{fs}$, and is much larger than the intrinsic lifetime of the system, e.g. $E_g^{-1}\sim 1\mathrm{fs}$. Therefore, one can think of energy dispersion as instantaneous energies of Hamiltonian being evolved adiabatically. Moreover, each band structure can be measured in the angle-resolved photoemission spectroscopy within the time domain \cite{hellmann2012time}. However, for a more precise evaluation of optical conductivity one may use the nonequilibrium two-time Green's function method \cite{Eckstein2008}.

The optical response to a pump laser pulse is rather complex in the BCS EI. The many-body peak is strongly pronounced at finite times long after the high energy pulse exposure and start to oscillate in time as shown in Fig.~\ref{nonopcosp}(b) through $\phi(t)$. Fig.~\ref{nonopcosp}(c) shows that at much higher laser pulse frequency, stronger manybody transition appears at finite times long after the pulse exposure.

\section{conclusions}\label{concolusion}
In this paper, we studied the exciton insulator phase in a one-dimensional chain of atomic $s-p$ orbitals in the presence of on-site inter-orbital Coulomb interaction. The model in the non-interacting regime presents a topological phase transition from a TI to trivial BI, providing a playground to study the possible formation of the exciton condensate in insulators with a nontrivial band topology. At equilibrium, our mean-field study reveals that contrary to the absence of exciton formation in the trivial BI phase, exciton condensate is formed in the TI phase at strong Coulomb interaction. This implies that the band inversion is crucial in the formation of exciton condensate, meaning that the in-plain and out-of-plain pseudomagnetic field related to bands with different parities compete. Our findings also show that BCS-BEC crossover is present in the low $s-p$ hybridization limit. Furthermore, we found that the collective modes in the phase direction are gaped as a consequence of inter-orbital coupling.

Motivated by recent pump-induced coherent dynamics of exciton condensates, we also studied the time-evolution of the exciton order parameter irradiated by a pump pulse and coupled to an optical phonon bath. We showed that exciton dynamics clearly depend upon the driven laser frequency, and the nature of EI phases, BCS vs BEC condensates, reveal different optical transitions. Also, the fingerprint of photo-induced bandgap shrinkage is observable as an energy redshift in the optical spectra of both EI types. Moreover, we found that long after the pulse exposure, the many-body transition is enhanced in the BSC EI phase, while it is suppressed in the BEC EI phase.  

The model studied in this paper can be generalized to include the spin degrees of freedom and possibly a richer phase diagram emegeres. Also, in the light of recent experiments on Ta$_2$NiSe$_5$ and its quasi one-dimensional nature, it would be interesting to see if the band parities might affect the exciton formation in this materisl, which we leave it for future study.

\section{ Acknowledgments}
The authors would like to acknowledge support from the Sharif University of Technology under Grant No. G960208 and Iran's National Elite Federation.

\appendix
\begin{figure*}[!htb]
    \center
    \includegraphics[width=\linewidth]{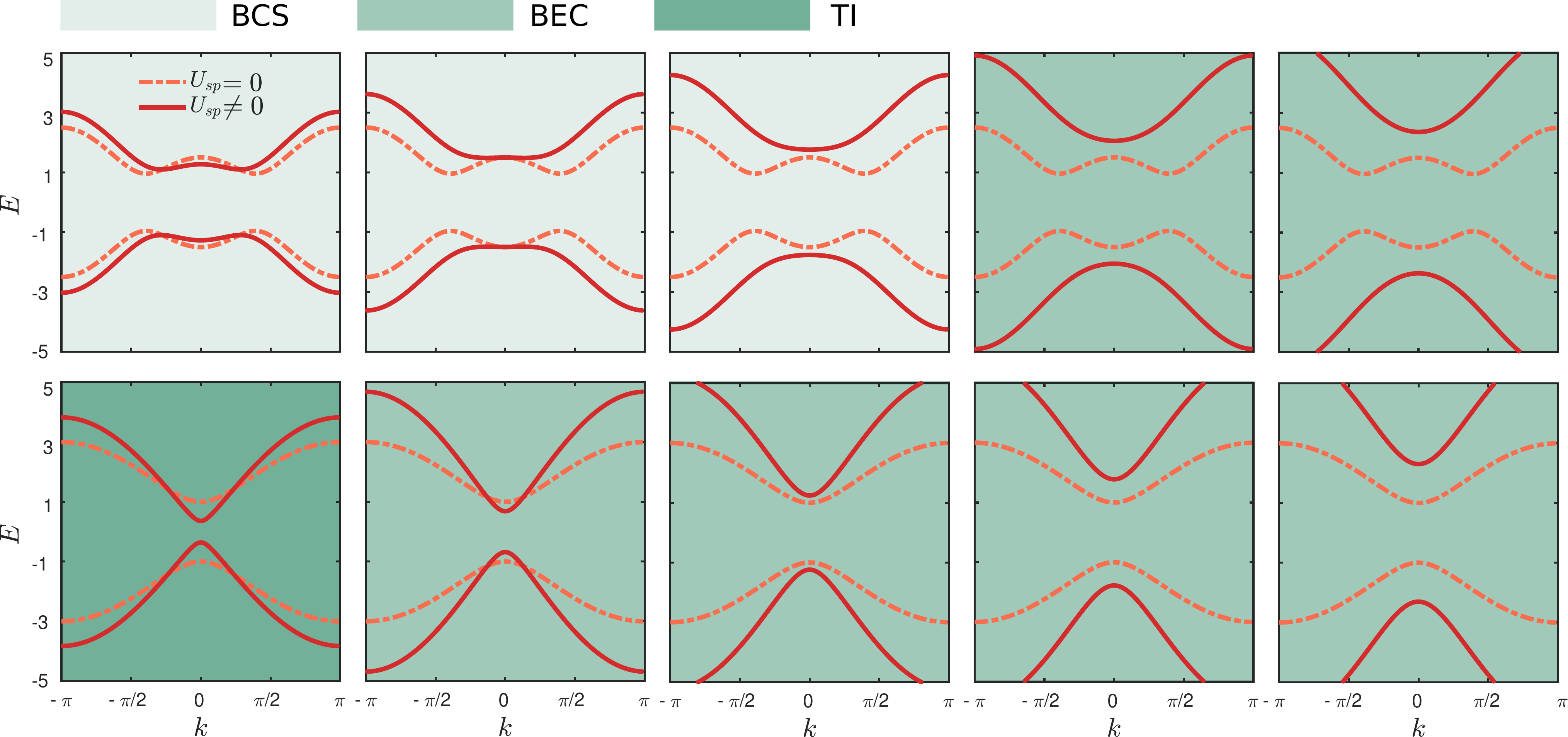}
    \caption{The evolution of band dispersion for a 1D $s-p$ chain with increment of $U_{sp}$ from $3$ to $7$. Top row plots are for a $s-p$ chain with $\epsilon=1/2, V_{sp}=1/2$, and the bottom row plots for one with $\epsilon=1/2, V_{sp}=1$. For bands being flattened near $k=0$ we use the increasing behavior of $\phi$ by $U_{sp}$ from BCS, bands with two minima and maxima, to characteize the crossover to the BEC phase. Here, we clearly see a BCS-BEC crossover with enhancement of inter-orbital Coulomb interaction $U_{sp}$.} \label{spbands2}
\end{figure*}

\section{Winding number}\label{windingsec}
In this appendix we present the details of computation of winding number for non-interacting Hamiltonian \eqref{nonintH}. The Bloch Hamiltonian of Eq.~\eqref{nonintH}, transforms under the following unitary transformation,
\begin{equation}\label{unitary}
U=\frac{1}{\sqrt{2}}\begin{bmatrix}
1 & 1 \\
1 & -1 
\end{bmatrix},
\end{equation}
to, 
\begin{equation}\label{Utrans}
H_0(k)=\left(\frac{\epsilon_s+\epsilon_p}{2}-t_s\cos k +t_p\cos k  \right)\mathbb{I}_{2\times 2}+\mathbf{d}(k)\cdot\boldsymbol{\sigma},
\end{equation}
where 
\begin{align}
&d_x(k)=\frac{\epsilon_s-\epsilon_p}{2}-t_s \cos k  -t_p \cos k  \\
&d_y(k)=2V_{sp}\sin k,~~~d_z(k)=0.
\end{align}

For $\epsilon_s=-\epsilon_p=\epsilon>0,~t_{s}=t_{p}=t=1$, the $H_0(k)$ is chiral symmetric, since $\{H_0(k),\sigma_z\}=0$. Thus, the winding number can be written as \cite{li2015winding}
{\allowdisplaybreaks
    \begin{align}\label{}
    \nu=\frac{1}{2i\pi}\int_{-\pi}^{\pi} dk \frac{d}{dk}\ln \tilde{d}(k)
    = \frac{\bar{V}}{2\pi}\int_{-\pi}^{\pi} dk \frac{\bar{\epsilon} \cos k -1}{(\bar{\epsilon}-\cos k  )^2+\bar{V}^2\sin^{2}k}
    \end{align}}
where we have used,
{\allowdisplaybreaks
    \begin{subequations}
        \begin{align}
        &\tilde{d}(k)=d_{x}(k)+id_{y}(k)=|\mathbf{d}(k)|e^{i\varphi_k},\\
        &|\mathbf{d}(k)|=\sqrt{(\epsilon-2t \cos k  )^2+4V^2_{sp}\sin^{2}k },\\
        &\varphi_{k}=~\tan^{-1}(d_y(k)/d_{x}(k)),\\
        &\bar{\epsilon}=~\epsilon/2t,\\
        &\bar{V}=~V_{sp}/t.
        \end{align}\end{subequations}}

Therefore, for a non-interacting $s-p$ chain ($U_{sp}=0$), when $\epsilon>2 \Rightarrow d_{x}(k)>0$, and the winding number becomes zero. In fact, in a high on-site energy regime, $s$ and $p$ orbitals are energetically separated leading to an effectively insignificant hybridization between $s$ and $p$ orbitals, thus a semiconductor forms. On the other hand, in a small $\epsilon$ regime, the $s-p$ hybridization significantly contributes to band inversion and leads to a TI phase forming.


\section{BCS-BEC crossover}\label{bcs-bec}
In the low $s-p$ hybridization energy limit, the band structure of a BCS type EI evolves as the inter-orbital Coulomb interaction is enhanced. The valence and conduction bands are flattened at $k=0$ with $U_{sp}$ increment until eventually, a BCS-BEC crossover occurs. Fig.~\ref{spbands2} illustrates the gradual evolution of band dispersion proportional to $U_{sp}$ strength.

\newpage

\begin{thebibliography}{39}%
\makeatletter
\providecommand \@ifxundefined [1]{%
 \@ifx{#1\undefined}
}%
\providecommand \@ifnum [1]{%
 \ifnum #1\expandafter \@firstoftwo
 \else \expandafter \@secondoftwo
 \fi
}%
\providecommand \@ifx [1]{%
 \ifx #1\expandafter \@firstoftwo
 \else \expandafter \@secondoftwo
 \fi
}%
\providecommand \natexlab [1]{#1}%
\providecommand \enquote  [1]{``#1''}%
\providecommand \bibnamefont  [1]{#1}%
\providecommand \bibfnamefont [1]{#1}%
\providecommand \citenamefont [1]{#1}%
\providecommand \href@noop [0]{\@secondoftwo}%
\providecommand \href [0]{\begingroup \@sanitize@url \@href}%
\providecommand \@href[1]{\@@startlink{#1}\@@href}%
\providecommand \@@href[1]{\endgroup#1\@@endlink}%
\providecommand \@sanitize@url [0]{\catcode `\\12\catcode `\$12\catcode
  `\&12\catcode `\#12\catcode `\^12\catcode `\_12\catcode `\%12\relax}%
\providecommand \@@startlink[1]{}%
\providecommand \@@endlink[0]{}%
\providecommand \url  [0]{\begingroup\@sanitize@url \@url }%
\providecommand \@url [1]{\endgroup\@href {#1}{\urlprefix }}%
\providecommand \urlprefix  [0]{URL }%
\providecommand \Eprint [0]{\href }%
\providecommand \doibase [0]{http://dx.doi.org/}%
\providecommand \selectlanguage [0]{\@gobble}%
\providecommand \bibinfo  [0]{\@secondoftwo}%
\providecommand \bibfield  [0]{\@secondoftwo}%
\providecommand \translation [1]{[#1]}%
\providecommand \BibitemOpen [0]{}%
\providecommand \bibitemStop [0]{}%
\providecommand \bibitemNoStop [0]{.\EOS\space}%
\providecommand \EOS [0]{\spacefactor3000\relax}%
\providecommand \BibitemShut  [1]{\csname bibitem#1\endcsname}%
\let\auto@bib@innerbib\@empty
\bibitem [{\citenamefont {Salvo}\ \emph {et~al.}(1986)\citenamefont {Salvo},
  \citenamefont {Chen}, \citenamefont {Fleming}, \citenamefont {Waszczak},
  \citenamefont {Dunn}, \citenamefont {Sunshine},\ and\ \citenamefont
  {Ibers}}]{Salvo1986physical}%
  \BibitemOpen
  \bibfield  {author} {\bibinfo {author} {\bibfnamefont {F.~D.}\ \bibnamefont
  {Salvo}}, \bibinfo {author} {\bibfnamefont {C.}~\bibnamefont {Chen}},
  \bibinfo {author} {\bibfnamefont {R.}~\bibnamefont {Fleming}}, \bibinfo
  {author} {\bibfnamefont {J.}~\bibnamefont {Waszczak}}, \bibinfo {author}
  {\bibfnamefont {R.}~\bibnamefont {Dunn}}, \bibinfo {author} {\bibfnamefont
  {S.}~\bibnamefont {Sunshine}}, \ and\ \bibinfo {author} {\bibfnamefont
  {J.~A.}\ \bibnamefont {Ibers}},\ }\href {\doibase
  https://doi.org/10.1016/0022-5088(86)90216-X} {\bibfield  {journal} {\bibinfo
   {journal} {J. Less- Common Met.}\ }\textbf {\bibinfo {volume} {116}},\
  \bibinfo {pages} {51} (\bibinfo {year} {1986})}\BibitemShut {NoStop}%
\bibitem [{\citenamefont {Wakisaka}\ \emph {et~al.}(2009)\citenamefont
  {Wakisaka}, \citenamefont {Sudayama}, \citenamefont {Takubo}, \citenamefont
  {Mizokawa}, \citenamefont {Arita}, \citenamefont {Namatame}, \citenamefont
  {Taniguchi}, \citenamefont {Katayama}, \citenamefont {Nohara},\ and\
  \citenamefont {Takagi}}]{Wakisaka2009}%
  \BibitemOpen
  \bibfield  {author} {\bibinfo {author} {\bibfnamefont {Y.}~\bibnamefont
  {Wakisaka}}, \bibinfo {author} {\bibfnamefont {T.}~\bibnamefont {Sudayama}},
  \bibinfo {author} {\bibfnamefont {K.}~\bibnamefont {Takubo}}, \bibinfo
  {author} {\bibfnamefont {T.}~\bibnamefont {Mizokawa}}, \bibinfo {author}
  {\bibfnamefont {M.}~\bibnamefont {Arita}}, \bibinfo {author} {\bibfnamefont
  {H.}~\bibnamefont {Namatame}}, \bibinfo {author} {\bibfnamefont
  {M.}~\bibnamefont {Taniguchi}}, \bibinfo {author} {\bibfnamefont
  {N.}~\bibnamefont {Katayama}}, \bibinfo {author} {\bibfnamefont
  {M.}~\bibnamefont {Nohara}}, \ and\ \bibinfo {author} {\bibfnamefont
  {H.}~\bibnamefont {Takagi}},\ }\href {\doibase
  10.1103/PhysRevLett.103.026402} {\bibfield  {journal} {\bibinfo  {journal}
  {Phys. Rev. Lett.}\ }\textbf {\bibinfo {volume} {103}},\ \bibinfo {pages}
  {026402} (\bibinfo {year} {2009})}\BibitemShut {NoStop}%
\bibitem [{\citenamefont {Hellmann}\ \emph {et~al.}(2012)\citenamefont
  {Hellmann}, \citenamefont {Rohwer}, \citenamefont {Kall{\"a}ne},
  \citenamefont {Hanff}, \citenamefont {Sohrt}, \citenamefont {Stange},
  \citenamefont {Carr}, \citenamefont {Murnane}, \citenamefont {Kapteyn},
  \citenamefont {Kipp} \emph {et~al.}}]{hellmann2012time}%
  \BibitemOpen
  \bibfield  {author} {\bibinfo {author} {\bibfnamefont {S.}~\bibnamefont
  {Hellmann}}, \bibinfo {author} {\bibfnamefont {T.}~\bibnamefont {Rohwer}},
  \bibinfo {author} {\bibfnamefont {M.}~\bibnamefont {Kall{\"a}ne}}, \bibinfo
  {author} {\bibfnamefont {K.}~\bibnamefont {Hanff}}, \bibinfo {author}
  {\bibfnamefont {C.}~\bibnamefont {Sohrt}}, \bibinfo {author} {\bibfnamefont
  {A.}~\bibnamefont {Stange}}, \bibinfo {author} {\bibfnamefont
  {A.}~\bibnamefont {Carr}}, \bibinfo {author} {\bibfnamefont {M.}~\bibnamefont
  {Murnane}}, \bibinfo {author} {\bibfnamefont {H.}~\bibnamefont {Kapteyn}},
  \bibinfo {author} {\bibfnamefont {L.}~\bibnamefont {Kipp}},  \emph {et~al.},\
  }\href {\doibase 10.1038/ncomms2078} {\bibfield  {journal} {\bibinfo
  {journal} {Nat. Commun}\ }\textbf {\bibinfo {volume} {3}},\ \bibinfo {pages}
  {1069} (\bibinfo {year} {2012})}\BibitemShut {NoStop}%
\bibitem [{\citenamefont {Zenker}\ \emph {et~al.}(2013)\citenamefont {Zenker},
  \citenamefont {Fehske}, \citenamefont {Beck}, \citenamefont {Monney},\ and\
  \citenamefont {Bishop}}]{Zenker2013chiral}%
  \BibitemOpen
  \bibfield  {author} {\bibinfo {author} {\bibfnamefont {B.}~\bibnamefont
  {Zenker}}, \bibinfo {author} {\bibfnamefont {H.}~\bibnamefont {Fehske}},
  \bibinfo {author} {\bibfnamefont {H.}~\bibnamefont {Beck}}, \bibinfo {author}
  {\bibfnamefont {C.}~\bibnamefont {Monney}}, \ and\ \bibinfo {author}
  {\bibfnamefont {A.~R.}\ \bibnamefont {Bishop}},\ }\href {\doibase
  10.1103/PhysRevB.88.075138} {\bibfield  {journal} {\bibinfo  {journal} {Phys.
  Rev. B}\ }\textbf {\bibinfo {volume} {88}},\ \bibinfo {pages} {075138}
  (\bibinfo {year} {2013})}\BibitemShut {NoStop}%
\bibitem [{\citenamefont {Kaneko}\ \emph {et~al.}(2015)\citenamefont {Kaneko},
  \citenamefont {Zenker}, \citenamefont {Fehske},\ and\ \citenamefont
  {Ohta}}]{Kaneko2015}%
  \BibitemOpen
  \bibfield  {author} {\bibinfo {author} {\bibfnamefont {T.}~\bibnamefont
  {Kaneko}}, \bibinfo {author} {\bibfnamefont {B.}~\bibnamefont {Zenker}},
  \bibinfo {author} {\bibfnamefont {H.}~\bibnamefont {Fehske}}, \ and\ \bibinfo
  {author} {\bibfnamefont {Y.}~\bibnamefont {Ohta}},\ }\href {\doibase
  10.1103/PhysRevB.92.115106} {\bibfield  {journal} {\bibinfo  {journal} {Phys.
  Rev. B}\ }\textbf {\bibinfo {volume} {92}},\ \bibinfo {pages} {115106}
  (\bibinfo {year} {2015})}\BibitemShut {NoStop}%
\bibitem [{\citenamefont {Larkin}\ \emph {et~al.}(2017)\citenamefont {Larkin},
  \citenamefont {Yaresko}, \citenamefont {Pr\"opper}, \citenamefont {Kikoin},
  \citenamefont {Lu}, \citenamefont {Takayama}, \citenamefont {Mathis},
  \citenamefont {Rost}, \citenamefont {Takagi}, \citenamefont {Keimer},\ and\
  \citenamefont {Boris}}]{Larkin2017Giant}%
  \BibitemOpen
  \bibfield  {author} {\bibinfo {author} {\bibfnamefont {T.~I.}\ \bibnamefont
  {Larkin}}, \bibinfo {author} {\bibfnamefont {A.~N.}\ \bibnamefont {Yaresko}},
  \bibinfo {author} {\bibfnamefont {D.}~\bibnamefont {Pr\"opper}}, \bibinfo
  {author} {\bibfnamefont {K.~A.}\ \bibnamefont {Kikoin}}, \bibinfo {author}
  {\bibfnamefont {Y.~F.}\ \bibnamefont {Lu}}, \bibinfo {author} {\bibfnamefont
  {T.}~\bibnamefont {Takayama}}, \bibinfo {author} {\bibfnamefont {Y.-L.}\
  \bibnamefont {Mathis}}, \bibinfo {author} {\bibfnamefont {A.~W.}\
  \bibnamefont {Rost}}, \bibinfo {author} {\bibfnamefont {H.}~\bibnamefont
  {Takagi}}, \bibinfo {author} {\bibfnamefont {B.}~\bibnamefont {Keimer}}, \
  and\ \bibinfo {author} {\bibfnamefont {A.~V.}\ \bibnamefont {Boris}},\ }\href
  {\doibase 10.1103/PhysRevB.95.195144} {\bibfield  {journal} {\bibinfo
  {journal} {Phys. Rev. B}\ }\textbf {\bibinfo {volume} {95}},\ \bibinfo
  {pages} {195144} (\bibinfo {year} {2017})}\BibitemShut {NoStop}%
\bibitem [{\citenamefont {Remez}\ and\ \citenamefont
  {Cooper}(2020)}]{Remez2020}%
  \BibitemOpen
  \bibfield  {author} {\bibinfo {author} {\bibfnamefont {B.}~\bibnamefont
  {Remez}}\ and\ \bibinfo {author} {\bibfnamefont {N.~R.}\ \bibnamefont
  {Cooper}},\ }\href {\doibase 10.1103/PhysRevB.101.235129} {\bibfield
  {journal} {\bibinfo  {journal} {Phys. Rev. B}\ }\textbf {\bibinfo {volume}
  {101}},\ \bibinfo {pages} {235129} (\bibinfo {year} {2020})}\BibitemShut
  {NoStop}%
\bibitem [{\citenamefont {Inayoshi}\ \emph {et~al.}(2020)\citenamefont
  {Inayoshi}, \citenamefont {Murakami},\ and\ \citenamefont
  {Koga}}]{Inayoshi2020}%
  \BibitemOpen
  \bibfield  {author} {\bibinfo {author} {\bibfnamefont {K.}~\bibnamefont
  {Inayoshi}}, \bibinfo {author} {\bibfnamefont {Y.}~\bibnamefont {Murakami}},
  \ and\ \bibinfo {author} {\bibfnamefont {A.}~\bibnamefont {Koga}},\ }\href
  {\doibase 10.7566/JPSJ.89.064002} {\bibfield  {journal} {\bibinfo  {journal}
  {Journal of the Physical Society of Japan}\ }\textbf {\bibinfo {volume}
  {89}},\ \bibinfo {pages} {064002} (\bibinfo {year} {2020})}\BibitemShut
  {NoStop}%
\bibitem [{\citenamefont {Kadosawa}\ \emph {et~al.}(2020)\citenamefont
  {Kadosawa}, \citenamefont {Nishimoto}, \citenamefont {Sugimoto},\ and\
  \citenamefont {Ohta}}]{kadosawa2020finite}%
  \BibitemOpen
  \bibfield  {author} {\bibinfo {author} {\bibfnamefont {M.}~\bibnamefont
  {Kadosawa}}, \bibinfo {author} {\bibfnamefont {S.}~\bibnamefont {Nishimoto}},
  \bibinfo {author} {\bibfnamefont {K.}~\bibnamefont {Sugimoto}}, \ and\
  \bibinfo {author} {\bibfnamefont {Y.}~\bibnamefont {Ohta}},\ }\href {\doibase
  10.7566/JPSJ.89.053706} {\bibfield  {journal} {\bibinfo  {journal} {Journal
  of the Physical Society of Japan}\ }\textbf {\bibinfo {volume} {89}},\
  \bibinfo {pages} {053706} (\bibinfo {year} {2020})}\BibitemShut {NoStop}%
\bibitem [{\citenamefont {Murakami}\ \emph
  {et~al.}(2020{\natexlab{a}})\citenamefont {Murakami}, \citenamefont
  {Sch{\"u}ler}, \citenamefont {Takayoshi},\ and\ \citenamefont
  {Werner}}]{murakami2020ultrafast}%
  \BibitemOpen
  \bibfield  {author} {\bibinfo {author} {\bibfnamefont {Y.}~\bibnamefont
  {Murakami}}, \bibinfo {author} {\bibfnamefont {M.}~\bibnamefont
  {Sch{\"u}ler}}, \bibinfo {author} {\bibfnamefont {S.}~\bibnamefont
  {Takayoshi}}, \ and\ \bibinfo {author} {\bibfnamefont {P.}~\bibnamefont
  {Werner}},\ }\href {\doibase 10.1103/PhysRevB.101.035203} {\bibfield
  {journal} {\bibinfo  {journal} {Phys. Rev. B}\ }\textbf {\bibinfo {volume}
  {101}},\ \bibinfo {pages} {035203} (\bibinfo {year}
  {2020}{\natexlab{a}})}\BibitemShut {NoStop}%
\bibitem [{\citenamefont {J{\'e}rome}\ \emph {et~al.}(1967)\citenamefont
  {J{\'e}rome}, \citenamefont {Rice},\ and\ \citenamefont
  {Kohn}}]{jerome1967excitonic}%
  \BibitemOpen
  \bibfield  {author} {\bibinfo {author} {\bibfnamefont {D.}~\bibnamefont
  {J{\'e}rome}}, \bibinfo {author} {\bibfnamefont {T.}~\bibnamefont {Rice}}, \
  and\ \bibinfo {author} {\bibfnamefont {W.}~\bibnamefont {Kohn}},\ }\href
  {\doibase 10.1103/PhysRev.158.462} {\bibfield  {journal} {\bibinfo  {journal}
  {Phys. Rev.}\ }\textbf {\bibinfo {volume} {158}},\ \bibinfo {pages} {462}
  (\bibinfo {year} {1967})}\BibitemShut {NoStop}%
\bibitem [{\citenamefont {Kohn}(1967)}]{kohn1967excitonic}%
  \BibitemOpen
  \bibfield  {author} {\bibinfo {author} {\bibfnamefont {W.}~\bibnamefont
  {Kohn}},\ }\href {\doibase 10.1103/PhysRevLett.19.439} {\bibfield  {journal}
  {\bibinfo  {journal} {Phys. Rev. Lett.}\ }\textbf {\bibinfo {volume} {19}},\
  \bibinfo {pages} {439} (\bibinfo {year} {1967})}\BibitemShut {NoStop}%
\bibitem [{\citenamefont {Halperin}\ and\ \citenamefont
  {Rice}(1968)}]{halperin1968possible}%
  \BibitemOpen
  \bibfield  {author} {\bibinfo {author} {\bibfnamefont {B.}~\bibnamefont
  {Halperin}}\ and\ \bibinfo {author} {\bibfnamefont {T.}~\bibnamefont
  {Rice}},\ }\href {\doibase 10.1103/RevModPhys.40.755} {\bibfield  {journal}
  {\bibinfo  {journal} {Rev. Mod. Phys.}\ }\textbf {\bibinfo {volume} {40}},\
  \bibinfo {pages} {755} (\bibinfo {year} {1968})}\BibitemShut {NoStop}%
\bibitem [{\citenamefont {Seki}\ \emph {et~al.}(2014)\citenamefont {Seki},
  \citenamefont {Wakisaka}, \citenamefont {Kaneko}, \citenamefont {Toriyama},
  \citenamefont {Konishi}, \citenamefont {Sudayama}, \citenamefont {Saini},
  \citenamefont {Arita}, \citenamefont {Namatame}, \citenamefont {Taniguchi},
  \citenamefont {Katayama}, \citenamefont {Nohara}, \citenamefont {Takagi},
  \citenamefont {Mizokawa},\ and\ \citenamefont {Ohta}}]{Seki2014Excitonic}%
  \BibitemOpen
  \bibfield  {author} {\bibinfo {author} {\bibfnamefont {K.}~\bibnamefont
  {Seki}}, \bibinfo {author} {\bibfnamefont {Y.}~\bibnamefont {Wakisaka}},
  \bibinfo {author} {\bibfnamefont {T.}~\bibnamefont {Kaneko}}, \bibinfo
  {author} {\bibfnamefont {T.}~\bibnamefont {Toriyama}}, \bibinfo {author}
  {\bibfnamefont {T.}~\bibnamefont {Konishi}}, \bibinfo {author} {\bibfnamefont
  {T.}~\bibnamefont {Sudayama}}, \bibinfo {author} {\bibfnamefont {N.~L.}\
  \bibnamefont {Saini}}, \bibinfo {author} {\bibfnamefont {M.}~\bibnamefont
  {Arita}}, \bibinfo {author} {\bibfnamefont {H.}~\bibnamefont {Namatame}},
  \bibinfo {author} {\bibfnamefont {M.}~\bibnamefont {Taniguchi}}, \bibinfo
  {author} {\bibfnamefont {N.}~\bibnamefont {Katayama}}, \bibinfo {author}
  {\bibfnamefont {M.}~\bibnamefont {Nohara}}, \bibinfo {author} {\bibfnamefont
  {H.}~\bibnamefont {Takagi}}, \bibinfo {author} {\bibfnamefont
  {T.}~\bibnamefont {Mizokawa}}, \ and\ \bibinfo {author} {\bibfnamefont
  {Y.}~\bibnamefont {Ohta}},\ }\href {\doibase 10.1103/PhysRevB.90.155116}
  {\bibfield  {journal} {\bibinfo  {journal} {Phys. Rev. B}\ }\textbf {\bibinfo
  {volume} {90}},\ \bibinfo {pages} {155116} (\bibinfo {year}
  {2014})}\BibitemShut {NoStop}%
\bibitem [{\citenamefont {Kim}\ \emph {et~al.}(2016)\citenamefont {Kim},
  \citenamefont {Kim}, \citenamefont {Kang}, \citenamefont {An}, \citenamefont
  {Kim}, \citenamefont {Eom}, \citenamefont {Lee}, \citenamefont {Park},
  \citenamefont {Kim}, \citenamefont {Choi} \emph {et~al.}}]{kim2016layer}%
  \BibitemOpen
  \bibfield  {author} {\bibinfo {author} {\bibfnamefont {S.~Y.}\ \bibnamefont
  {Kim}}, \bibinfo {author} {\bibfnamefont {Y.}~\bibnamefont {Kim}}, \bibinfo
  {author} {\bibfnamefont {C.-J.}\ \bibnamefont {Kang}}, \bibinfo {author}
  {\bibfnamefont {E.-S.}\ \bibnamefont {An}}, \bibinfo {author} {\bibfnamefont
  {H.~K.}\ \bibnamefont {Kim}}, \bibinfo {author} {\bibfnamefont {M.~J.}\
  \bibnamefont {Eom}}, \bibinfo {author} {\bibfnamefont {M.}~\bibnamefont
  {Lee}}, \bibinfo {author} {\bibfnamefont {C.}~\bibnamefont {Park}}, \bibinfo
  {author} {\bibfnamefont {T.-H.}\ \bibnamefont {Kim}}, \bibinfo {author}
  {\bibfnamefont {H.~C.}\ \bibnamefont {Choi}},  \emph {et~al.},\ }\href
  {\doibase 10.1021/acsnano.6b04796} {\bibfield  {journal} {\bibinfo  {journal}
  {ACS nano}\ }\textbf {\bibinfo {volume} {10}},\ \bibinfo {pages} {8888}
  (\bibinfo {year} {2016})}\BibitemShut {NoStop}%
\bibitem [{\citenamefont {Yu}\ \emph {et~al.}(2014)\citenamefont {Yu},
  \citenamefont {Liu}, \citenamefont {Gong}, \citenamefont {Xu},\ and\
  \citenamefont {Yao}}]{Yu2014Dirac}%
  \BibitemOpen
  \bibfield  {author} {\bibinfo {author} {\bibfnamefont {H.}~\bibnamefont
  {Yu}}, \bibinfo {author} {\bibfnamefont {G.~B.}\ \bibnamefont {Liu}},
  \bibinfo {author} {\bibfnamefont {P.}~\bibnamefont {Gong}}, \bibinfo {author}
  {\bibfnamefont {X.}~\bibnamefont {Xu}}, \ and\ \bibinfo {author}
  {\bibfnamefont {W.}~\bibnamefont {Yao}},\ }\href {\doibase
  10.1038/ncomms4876} {\bibfield  {journal} {\bibinfo  {journal} {Nat.
  Commun.}\ }\textbf {\bibinfo {volume} {5}},\ \bibinfo {pages} {1} (\bibinfo
  {year} {2014})}\BibitemShut {NoStop}%
\bibitem [{\citenamefont {Zhou}\ \emph {et~al.}(2015)\citenamefont {Zhou},
  \citenamefont {Shan}, \citenamefont {Yao},\ and\ \citenamefont
  {Xiao}}]{Zhou2015berry}%
  \BibitemOpen
  \bibfield  {author} {\bibinfo {author} {\bibfnamefont {J.}~\bibnamefont
  {Zhou}}, \bibinfo {author} {\bibfnamefont {W.~Y.}\ \bibnamefont {Shan}},
  \bibinfo {author} {\bibfnamefont {W.}~\bibnamefont {Yao}}, \ and\ \bibinfo
  {author} {\bibfnamefont {D.}~\bibnamefont {Xiao}},\ }\href {\doibase
  10.1103/PhysRevLett.115.166803} {\bibfield  {journal} {\bibinfo  {journal}
  {Phys. Rev. Lett.}\ }\textbf {\bibinfo {volume} {115}},\ \bibinfo {pages}
  {166803} (\bibinfo {year} {2015})}\BibitemShut {NoStop}%
\bibitem [{\citenamefont {Srivastava}\ and\ \citenamefont
  {Imamo\ifmmode~\breve{g}\else \u{g}\fi{}lu}(2015)}]{Srivastava2015}%
  \BibitemOpen
  \bibfield  {author} {\bibinfo {author} {\bibfnamefont {A.}~\bibnamefont
  {Srivastava}}\ and\ \bibinfo {author} {\bibfnamefont {A.~m.~c.}\ \bibnamefont
  {Imamo\ifmmode~\breve{g}\else \u{g}\fi{}lu}},\ }\href {\doibase
  10.1103/PhysRevLett.115.166802} {\bibfield  {journal} {\bibinfo  {journal}
  {Phys. Rev. Lett.}\ }\textbf {\bibinfo {volume} {115}},\ \bibinfo {pages}
  {166802} (\bibinfo {year} {2015})}\BibitemShut {NoStop}%
\bibitem [{\citenamefont {Jin}\ \emph {et~al.}(2019)\citenamefont {Jin},
  \citenamefont {Regan}, \citenamefont {Yan}, \citenamefont {Utama},
  \citenamefont {Wang}, \citenamefont {Zhao}, \citenamefont {Qin},
  \citenamefont {Yang}, \citenamefont {Zheng}, \citenamefont {Shi} \emph
  {et~al.}}]{jin2019observation}%
  \BibitemOpen
  \bibfield  {author} {\bibinfo {author} {\bibfnamefont {C.}~\bibnamefont
  {Jin}}, \bibinfo {author} {\bibfnamefont {E.~C.}\ \bibnamefont {Regan}},
  \bibinfo {author} {\bibfnamefont {A.}~\bibnamefont {Yan}}, \bibinfo {author}
  {\bibfnamefont {M.~I.~B.}\ \bibnamefont {Utama}}, \bibinfo {author}
  {\bibfnamefont {D.}~\bibnamefont {Wang}}, \bibinfo {author} {\bibfnamefont
  {S.}~\bibnamefont {Zhao}}, \bibinfo {author} {\bibfnamefont {Y.}~\bibnamefont
  {Qin}}, \bibinfo {author} {\bibfnamefont {S.}~\bibnamefont {Yang}}, \bibinfo
  {author} {\bibfnamefont {Z.}~\bibnamefont {Zheng}}, \bibinfo {author}
  {\bibfnamefont {S.}~\bibnamefont {Shi}},  \emph {et~al.},\ }\href {\doibase
  10.1038/s41586-019-0976-y} {\bibfield  {journal} {\bibinfo  {journal}
  {Nature}\ }\textbf {\bibinfo {volume} {567}},\ \bibinfo {pages} {76}
  (\bibinfo {year} {2019})}\BibitemShut {NoStop}%
\bibitem [{\citenamefont {Fang}\ \emph {et~al.}(2019)\citenamefont {Fang},
  \citenamefont {Han}, \citenamefont {Robert}, \citenamefont {Semina},
  \citenamefont {Lagarde}, \citenamefont {Courtade}, \citenamefont {Taniguchi},
  \citenamefont {Watanabe}, \citenamefont {Amand}, \citenamefont {Urbaszek}
  \emph {et~al.}}]{fang2019control}%
  \BibitemOpen
  \bibfield  {author} {\bibinfo {author} {\bibfnamefont {H.}~\bibnamefont
  {Fang}}, \bibinfo {author} {\bibfnamefont {B.}~\bibnamefont {Han}}, \bibinfo
  {author} {\bibfnamefont {C.}~\bibnamefont {Robert}}, \bibinfo {author}
  {\bibfnamefont {M.}~\bibnamefont {Semina}}, \bibinfo {author} {\bibfnamefont
  {D.}~\bibnamefont {Lagarde}}, \bibinfo {author} {\bibfnamefont
  {E.}~\bibnamefont {Courtade}}, \bibinfo {author} {\bibfnamefont
  {T.}~\bibnamefont {Taniguchi}}, \bibinfo {author} {\bibfnamefont
  {K.}~\bibnamefont {Watanabe}}, \bibinfo {author} {\bibfnamefont
  {T.}~\bibnamefont {Amand}}, \bibinfo {author} {\bibfnamefont
  {B.}~\bibnamefont {Urbaszek}},  \emph {et~al.},\ }\href {\doibase
  10.1103/PhysRevLett.123.067401} {\bibfield  {journal} {\bibinfo  {journal}
  {Phys. Rev. Lett.}\ }\textbf {\bibinfo {volume} {123}},\ \bibinfo {pages}
  {067401} (\bibinfo {year} {2019})}\BibitemShut {NoStop}%
\bibitem [{\citenamefont {Cadiz}\ \emph {et~al.}(2017)\citenamefont {Cadiz},
  \citenamefont {Courtade}, \citenamefont {Robert}, \citenamefont {Wang},
  \citenamefont {Shen}, \citenamefont {Cai}, \citenamefont {Taniguchi},
  \citenamefont {Watanabe}, \citenamefont {Carrere}, \citenamefont {Lagarde}
  \emph {et~al.}}]{cadiz2017excitonic}%
  \BibitemOpen
  \bibfield  {author} {\bibinfo {author} {\bibfnamefont {F.}~\bibnamefont
  {Cadiz}}, \bibinfo {author} {\bibfnamefont {E.}~\bibnamefont {Courtade}},
  \bibinfo {author} {\bibfnamefont {C.}~\bibnamefont {Robert}}, \bibinfo
  {author} {\bibfnamefont {G.}~\bibnamefont {Wang}}, \bibinfo {author}
  {\bibfnamefont {Y.}~\bibnamefont {Shen}}, \bibinfo {author} {\bibfnamefont
  {H.}~\bibnamefont {Cai}}, \bibinfo {author} {\bibfnamefont {T.}~\bibnamefont
  {Taniguchi}}, \bibinfo {author} {\bibfnamefont {K.}~\bibnamefont {Watanabe}},
  \bibinfo {author} {\bibfnamefont {H.}~\bibnamefont {Carrere}}, \bibinfo
  {author} {\bibfnamefont {D.}~\bibnamefont {Lagarde}},  \emph {et~al.},\
  }\href {\doibase 10.1103/PhysRevX.7.021026} {\bibfield  {journal} {\bibinfo
  {journal} {Phys. Rev. X}\ }\textbf {\bibinfo {volume} {7}},\ \bibinfo {pages}
  {021026} (\bibinfo {year} {2017})}\BibitemShut {NoStop}%
\bibitem [{\citenamefont {Lu}\ \emph {et~al.}(2017)\citenamefont {Lu},
  \citenamefont {Kono}, \citenamefont {Larkin}, \citenamefont {Rost},
  \citenamefont {Takayama}, \citenamefont {Boris}, \citenamefont {Keimer},\
  and\ \citenamefont {Takagi}}]{lu2017zero}%
  \BibitemOpen
  \bibfield  {author} {\bibinfo {author} {\bibfnamefont {Y.}~\bibnamefont
  {Lu}}, \bibinfo {author} {\bibfnamefont {H.}~\bibnamefont {Kono}}, \bibinfo
  {author} {\bibfnamefont {T.}~\bibnamefont {Larkin}}, \bibinfo {author}
  {\bibfnamefont {A.}~\bibnamefont {Rost}}, \bibinfo {author} {\bibfnamefont
  {T.}~\bibnamefont {Takayama}}, \bibinfo {author} {\bibfnamefont
  {A.}~\bibnamefont {Boris}}, \bibinfo {author} {\bibfnamefont
  {B.}~\bibnamefont {Keimer}}, \ and\ \bibinfo {author} {\bibfnamefont
  {H.}~\bibnamefont {Takagi}},\ }\href {\doibase 10.1038/ncomms14408}
  {\bibfield  {journal} {\bibinfo  {journal} {Nat. Commun}\ }\textbf {\bibinfo
  {volume} {8}},\ \bibinfo {pages} {14408} (\bibinfo {year}
  {2017})}\BibitemShut {NoStop}%
\bibitem [{\citenamefont {Werdehausen}\ \emph {et~al.}(2018)\citenamefont
  {Werdehausen}, \citenamefont {Takayama}, \citenamefont {H{\"o}ppner},
  \citenamefont {Albrecht}, \citenamefont {Rost}, \citenamefont {Lu},
  \citenamefont {Manske}, \citenamefont {Takagi},\ and\ \citenamefont
  {Kaiser}}]{Werdehausen2018}%
  \BibitemOpen
  \bibfield  {author} {\bibinfo {author} {\bibfnamefont {D.}~\bibnamefont
  {Werdehausen}}, \bibinfo {author} {\bibfnamefont {T.}~\bibnamefont
  {Takayama}}, \bibinfo {author} {\bibfnamefont {M.}~\bibnamefont
  {H{\"o}ppner}}, \bibinfo {author} {\bibfnamefont {G.}~\bibnamefont
  {Albrecht}}, \bibinfo {author} {\bibfnamefont {A.~W.}\ \bibnamefont {Rost}},
  \bibinfo {author} {\bibfnamefont {Y.}~\bibnamefont {Lu}}, \bibinfo {author}
  {\bibfnamefont {D.}~\bibnamefont {Manske}}, \bibinfo {author} {\bibfnamefont
  {H.}~\bibnamefont {Takagi}}, \ and\ \bibinfo {author} {\bibfnamefont
  {S.}~\bibnamefont {Kaiser}},\ }\href {\doibase 10.1126/sciadv.aap8652}
  {\bibfield  {journal} {\bibinfo  {journal} {Sci. Adv.}\ }\textbf {\bibinfo
  {volume} {4}},\ \bibinfo {pages} {eaap8652} (\bibinfo {year}
  {2018})}\BibitemShut {NoStop}%
\bibitem [{\citenamefont {Mazza}\ \emph {et~al.}(2019)\citenamefont {Mazza},
  \citenamefont {R{\"{o}}sner}, \citenamefont {Windg{\"{a}}tter}, \citenamefont
  {Latini}, \citenamefont {H{\"{u}}bener}, \citenamefont {Millis},
  \citenamefont {Rubio},\ and\ \citenamefont {Georges}}]{Mazza2019}%
  \BibitemOpen
  \bibfield  {author} {\bibinfo {author} {\bibfnamefont {G.}~\bibnamefont
  {Mazza}}, \bibinfo {author} {\bibfnamefont {M.}~\bibnamefont {R{\"{o}}sner}},
  \bibinfo {author} {\bibfnamefont {L.}~\bibnamefont {Windg{\"{a}}tter}},
  \bibinfo {author} {\bibfnamefont {S.}~\bibnamefont {Latini}}, \bibinfo
  {author} {\bibfnamefont {H.}~\bibnamefont {H{\"{u}}bener}}, \bibinfo {author}
  {\bibfnamefont {A.~J.}\ \bibnamefont {Millis}}, \bibinfo {author}
  {\bibfnamefont {A.}~\bibnamefont {Rubio}}, \ and\ \bibinfo {author}
  {\bibfnamefont {A.}~\bibnamefont {Georges}},\ }\href {\doibase
  10.1103/PhysRevLett.124.197601} {\bibfield  {journal} {\bibinfo  {journal}
  {Phys. Rev. Lett.}\ }\textbf {\bibinfo {volume} {124}},\ \bibinfo {pages}
  {197601} (\bibinfo {year} {2019})}\BibitemShut {NoStop}%
\bibitem [{\citenamefont {Tang}\ \emph {et~al.}()\citenamefont {Tang},
  \citenamefont {Wang}, \citenamefont {Duan}, \citenamefont {Yang},
  \citenamefont {Huang}, \citenamefont {Guo}, \citenamefont {Qian},\ and\
  \citenamefont {Zhang}}]{Tang2020}%
  \BibitemOpen
  \bibfield  {author} {\bibinfo {author} {\bibfnamefont {T.}~\bibnamefont
  {Tang}}, \bibinfo {author} {\bibfnamefont {H.}~\bibnamefont {Wang}}, \bibinfo
  {author} {\bibfnamefont {S.}~\bibnamefont {Duan}}, \bibinfo {author}
  {\bibfnamefont {Y.}~\bibnamefont {Yang}}, \bibinfo {author} {\bibfnamefont
  {C.}~\bibnamefont {Huang}}, \bibinfo {author} {\bibfnamefont
  {Y.}~\bibnamefont {Guo}}, \bibinfo {author} {\bibfnamefont {D.}~\bibnamefont
  {Qian}}, \ and\ \bibinfo {author} {\bibfnamefont {W.}~\bibnamefont {Zhang}},\
  }\href@noop {} {\bibinfo  {journal} {arXiv preprint
  \href{https://arxiv.org/abs/2003.00514}{arXiv:2003.00514 (2020)}}\
  }\BibitemShut {NoStop}%
\bibitem [{\citenamefont {Andrich}\ \emph {et~al.}()\citenamefont {Andrich},
  \citenamefont {Bretscher}, \citenamefont {Murakami}, \citenamefont
  {Gole{\v{z}}}, \citenamefont {Remez}, \citenamefont {Telang}, \citenamefont
  {Singh}, \citenamefont {Harnagea}, \citenamefont {Cooper}, \citenamefont
  {Millis} \emph {et~al.}}]{andrich2020imaging}%
  \BibitemOpen
\bibfield  {journal} {  }\bibfield  {author} {\bibinfo {author} {\bibfnamefont
  {P.}~\bibnamefont {Andrich}}, \bibinfo {author} {\bibfnamefont {H.~M.}\
  \bibnamefont {Bretscher}}, \bibinfo {author} {\bibfnamefont {Y.}~\bibnamefont
  {Murakami}}, \bibinfo {author} {\bibfnamefont {D.}~\bibnamefont
  {Gole{\v{z}}}}, \bibinfo {author} {\bibfnamefont {B.}~\bibnamefont {Remez}},
  \bibinfo {author} {\bibfnamefont {P.}~\bibnamefont {Telang}}, \bibinfo
  {author} {\bibfnamefont {A.}~\bibnamefont {Singh}}, \bibinfo {author}
  {\bibfnamefont {L.}~\bibnamefont {Harnagea}}, \bibinfo {author}
  {\bibfnamefont {N.~R.}\ \bibnamefont {Cooper}}, \bibinfo {author}
  {\bibfnamefont {A.~J.}\ \bibnamefont {Millis}},  \emph {et~al.},\ }\href@noop
  {} {\bibinfo  {journal} {arXiv preprint
  \href{https://arxiv.org/abs/2003.10799}{arXiv:2003.10799 (2020)}}\
  }\BibitemShut {NoStop}%
\bibitem [{\citenamefont {Sugimoto}\ \emph {et~al.}(2018)\citenamefont
  {Sugimoto}, \citenamefont {Nishimoto}, \citenamefont {Kaneko},\ and\
  \citenamefont {Ohta}}]{Sugimoto2018strong}%
  \BibitemOpen
\bibfield  {journal} {  }\bibfield  {author} {\bibinfo {author} {\bibfnamefont
  {K.}~\bibnamefont {Sugimoto}}, \bibinfo {author} {\bibfnamefont
  {S.}~\bibnamefont {Nishimoto}}, \bibinfo {author} {\bibfnamefont
  {T.}~\bibnamefont {Kaneko}}, \ and\ \bibinfo {author} {\bibfnamefont
  {Y.}~\bibnamefont {Ohta}},\ }\href {\doibase 10.1103/PhysRevLett.120.247602}
  {\bibfield  {journal} {\bibinfo  {journal} {Phys. Rev. Lett.}\ }\textbf
  {\bibinfo {volume} {120}},\ \bibinfo {pages} {247602} (\bibinfo {year}
  {2018})}\BibitemShut {NoStop}%
\bibitem [{\citenamefont {Baldini}\ \emph {et~al.}()\citenamefont {Baldini},
  \citenamefont {Zong}, \citenamefont {Choi}, \citenamefont {Lee},
  \citenamefont {Michael}, \citenamefont {Windgaetter}, \citenamefont {Mazin},
  \citenamefont {Latini} \emph {et~al.}}]{Baldini2020spontaneous}%
  \BibitemOpen
  \bibfield  {author} {\bibinfo {author} {\bibfnamefont {E.}~\bibnamefont
  {Baldini}}, \bibinfo {author} {\bibfnamefont {A.}~\bibnamefont {Zong}},
  \bibinfo {author} {\bibfnamefont {D.}~\bibnamefont {Choi}}, \bibinfo {author}
  {\bibfnamefont {C.}~\bibnamefont {Lee}}, \bibinfo {author} {\bibfnamefont
  {M.~H.}\ \bibnamefont {Michael}}, \bibinfo {author} {\bibfnamefont
  {L.}~\bibnamefont {Windgaetter}}, \bibinfo {author} {\bibfnamefont {I.~I.}\
  \bibnamefont {Mazin}}, \bibinfo {author} {\bibfnamefont {S.}~\bibnamefont
  {Latini}},  \emph {et~al.},\ }\href@noop {} {\bibinfo  {journal} {arXiv
  preprint \href{https://arxiv.org/abs/2007.02909}{arXiv:2007.02909 (2020)}}\
  }\BibitemShut {NoStop}%
\bibitem [{\citenamefont {Kim}\ \emph {et~al.}()\citenamefont {Kim},
  \citenamefont {Schulz}, \citenamefont {Takayama}, \citenamefont {Isobe},
  \citenamefont {Takagi},\ and\ \citenamefont {Kaiser}}]{Kim2020Observation}%
  \BibitemOpen
\bibfield  {journal} {  }\bibfield  {author} {\bibinfo {author} {\bibfnamefont
  {M.-J.}\ \bibnamefont {Kim}}, \bibinfo {author} {\bibfnamefont
  {A.}~\bibnamefont {Schulz}}, \bibinfo {author} {\bibfnamefont
  {T.}~\bibnamefont {Takayama}}, \bibinfo {author} {\bibfnamefont
  {M.}~\bibnamefont {Isobe}}, \bibinfo {author} {\bibfnamefont
  {H.}~\bibnamefont {Takagi}}, \ and\ \bibinfo {author} {\bibfnamefont
  {S.}~\bibnamefont {Kaiser}},\ }\href@noop {} {\bibinfo  {journal} {arXiv
  preprint \href{http://arxiv.org/abs/2007.01723}{arXiv:2007.01723 (2020)}}\
  }\BibitemShut {NoStop}%
\bibitem [{\citenamefont {Zenker}\ \emph {et~al.}(2010)\citenamefont {Zenker},
  \citenamefont {Ihle}, \citenamefont {Bronold},\ and\ \citenamefont
  {Fehske}}]{Zenker2010}%
  \BibitemOpen
\bibfield  {journal} {  }\bibfield  {author} {\bibinfo {author} {\bibfnamefont
  {B.}~\bibnamefont {Zenker}}, \bibinfo {author} {\bibfnamefont
  {D.}~\bibnamefont {Ihle}}, \bibinfo {author} {\bibfnamefont {F.~X.}\
  \bibnamefont {Bronold}}, \ and\ \bibinfo {author} {\bibfnamefont
  {H.}~\bibnamefont {Fehske}},\ }\href {\doibase 10.1103/PhysRevB.81.115122}
  {\bibfield  {journal} {\bibinfo  {journal} {Phys. Rev. B}\ }\textbf {\bibinfo
  {volume} {81}},\ \bibinfo {pages} {115122} (\bibinfo {year}
  {2010})}\BibitemShut {NoStop}%
\bibitem [{\citenamefont {Shockley}(1939)}]{shockley1939surface}%
  \BibitemOpen
  \bibfield  {author} {\bibinfo {author} {\bibfnamefont {W.}~\bibnamefont
  {Shockley}},\ }\href {\doibase 10.1103/PhysRev.56.317} {\bibfield  {journal}
  {\bibinfo  {journal} {Phys. Rev.}\ }\textbf {\bibinfo {volume} {56}},\
  \bibinfo {pages} {317} (\bibinfo {year} {1939})}\BibitemShut {NoStop}%
\bibitem [{\citenamefont {Continentino}\ \emph {et~al.}(2014)\citenamefont
  {Continentino}, \citenamefont {Caldas}, \citenamefont {Nozadze},\ and\
  \citenamefont {Trivedi}}]{continentino2014topological}%
  \BibitemOpen
  \bibfield  {author} {\bibinfo {author} {\bibfnamefont {M.~A.}\ \bibnamefont
  {Continentino}}, \bibinfo {author} {\bibfnamefont {H.}~\bibnamefont
  {Caldas}}, \bibinfo {author} {\bibfnamefont {D.}~\bibnamefont {Nozadze}}, \
  and\ \bibinfo {author} {\bibfnamefont {N.}~\bibnamefont {Trivedi}},\ }\href
  {\doibase 10.1016/j.physleta.2014.09.022} {\bibfield  {journal} {\bibinfo
  {journal} {Phys. Lett. A}\ }\textbf {\bibinfo {volume} {378}},\ \bibinfo
  {pages} {3340} (\bibinfo {year} {2014})}\BibitemShut {NoStop}%
\bibitem [{\citenamefont {Kune{\v{s}}}(2015)}]{kunevs2015excitonic}%
  \BibitemOpen
  \bibfield  {author} {\bibinfo {author} {\bibfnamefont {J.}~\bibnamefont
  {Kune{\v{s}}}},\ }\href {\doibase 10.1088/0953-8984/27/33/333201} {\bibfield
  {journal} {\bibinfo  {journal} {J. Phys. Condens. Matter}\ }\textbf {\bibinfo
  {volume} {27}},\ \bibinfo {pages} {333201} (\bibinfo {year}
  {2015})}\BibitemShut {NoStop}%
\bibitem [{\citenamefont {Murakami}\ \emph {et~al.}(2017)\citenamefont
  {Murakami}, \citenamefont {Gole{\v{z}}}, \citenamefont {Eckstein},\ and\
  \citenamefont {Werner}}]{murakami2017photoinduced}%
  \BibitemOpen
  \bibfield  {author} {\bibinfo {author} {\bibfnamefont {Y.}~\bibnamefont
  {Murakami}}, \bibinfo {author} {\bibfnamefont {D.}~\bibnamefont
  {Gole{\v{z}}}}, \bibinfo {author} {\bibfnamefont {M.}~\bibnamefont
  {Eckstein}}, \ and\ \bibinfo {author} {\bibfnamefont {P.}~\bibnamefont
  {Werner}},\ }\href {\doibase 10.1103/PhysRevLett.119.247601} {\bibfield
  {journal} {\bibinfo  {journal} {Phys. Rev. Lett.}\ }\textbf {\bibinfo
  {volume} {119}},\ \bibinfo {pages} {247601} (\bibinfo {year}
  {2017})}\BibitemShut {NoStop}%
\bibitem [{\citenamefont {Tanabe}\ \emph {et~al.}(2018)\citenamefont {Tanabe},
  \citenamefont {Sugimoto},\ and\ \citenamefont
  {Ohta}}]{tanabe2018nonequilibrium}%
  \BibitemOpen
  \bibfield  {author} {\bibinfo {author} {\bibfnamefont {T.}~\bibnamefont
  {Tanabe}}, \bibinfo {author} {\bibfnamefont {K.}~\bibnamefont {Sugimoto}}, \
  and\ \bibinfo {author} {\bibfnamefont {Y.}~\bibnamefont {Ohta}},\ }\href
  {\doibase 10.1103/PhysRevB.98.235127} {\bibfield  {journal} {\bibinfo
  {journal} {Phys. Rev. B}\ }\textbf {\bibinfo {volume} {98}},\ \bibinfo
  {pages} {235127} (\bibinfo {year} {2018})}\BibitemShut {NoStop}%
\bibitem [{\citenamefont {Murakami}\ \emph
  {et~al.}(2020{\natexlab{b}})\citenamefont {Murakami}, \citenamefont
  {Gole\ifmmode~\check{z}\else \v{z}\fi{}}, \citenamefont {Kaneko},
  \citenamefont {Koga}, \citenamefont {Millis},\ and\ \citenamefont
  {Werner}}]{Murakami2020Collective}%
  \BibitemOpen
  \bibfield  {author} {\bibinfo {author} {\bibfnamefont {Y.}~\bibnamefont
  {Murakami}}, \bibinfo {author} {\bibfnamefont {D.}~\bibnamefont
  {Gole\ifmmode~\check{z}\else \v{z}\fi{}}}, \bibinfo {author} {\bibfnamefont
  {T.}~\bibnamefont {Kaneko}}, \bibinfo {author} {\bibfnamefont
  {A.}~\bibnamefont {Koga}}, \bibinfo {author} {\bibfnamefont {A.~J.}\
  \bibnamefont {Millis}}, \ and\ \bibinfo {author} {\bibfnamefont
  {P.}~\bibnamefont {Werner}},\ }\href {\doibase 10.1103/PhysRevB.101.195118}
  {\bibfield  {journal} {\bibinfo  {journal} {Phys. Rev. B}\ }\textbf {\bibinfo
  {volume} {101}},\ \bibinfo {pages} {195118} (\bibinfo {year}
  {2020}{\natexlab{b}})}\BibitemShut {NoStop}%
\bibitem [{\citenamefont {Nakahara}(2003)}]{Nakahara2003}%
  \BibitemOpen
  \bibfield  {author} {\bibinfo {author} {\bibfnamefont {M.}~\bibnamefont
  {Nakahara}},\ }\href@noop {} {\emph {\bibinfo {title} {Geometry, Topology and
  Physics}}}\ (\bibinfo  {publisher} {CRC Press},\ \bibinfo {year}
  {2003})\BibitemShut {NoStop}%
\bibitem [{\citenamefont {Kaushal}\ \emph {et~al.}()\citenamefont {Kaushal},
  \citenamefont {Soni}, \citenamefont {Nocera}, \citenamefont {Alvarez},\ and\
  \citenamefont {Dagotto}}]{kaushal2020bcs}%
  \BibitemOpen
  \bibfield  {author} {\bibinfo {author} {\bibfnamefont {N.}~\bibnamefont
  {Kaushal}}, \bibinfo {author} {\bibfnamefont {R.}~\bibnamefont {Soni}},
  \bibinfo {author} {\bibfnamefont {A.}~\bibnamefont {Nocera}}, \bibinfo
  {author} {\bibfnamefont {G.}~\bibnamefont {Alvarez}}, \ and\ \bibinfo
  {author} {\bibfnamefont {E.}~\bibnamefont {Dagotto}},\ }\href@noop {}
  {\bibinfo  {journal} {arXiv preprint
  \href{https://arxiv.org/abs/2002.07351}{arXiv:2002.07351 (2020)}}\
  }\BibitemShut {NoStop}%
\bibitem [{\citenamefont {Li}\ \emph {et~al.}(2015)\citenamefont {Li},
  \citenamefont {Yang},\ and\ \citenamefont {Chen}}]{li2015winding}%
  \BibitemOpen
\bibfield  {journal} {  }\bibfield  {author} {\bibinfo {author} {\bibfnamefont
  {L.}~\bibnamefont {Li}}, \bibinfo {author} {\bibfnamefont {C.}~\bibnamefont
  {Yang}}, \ and\ \bibinfo {author} {\bibfnamefont {S.}~\bibnamefont {Chen}},\
  }\href {\doibase 10.1209/0295-5075/112/10004} {\bibfield  {journal} {\bibinfo
   {journal} {EPL}\ }\textbf {\bibinfo {volume} {112}},\ \bibinfo {pages}
  {10004} (\bibinfo {year} {2015})}\BibitemShut {NoStop}%
\bibitem [{\citenamefont {Tanaka}\ \emph {et~al.}(2018)\citenamefont {Tanaka},
	\citenamefont {Daira},\ and\ \citenamefont {Yonemitsu}}]{Tanaka2018}%
\BibitemOpen
\bibfield  {author} {\bibinfo {author} {\bibfnamefont {Y.}~\bibnamefont
		{Tanaka}}, \bibinfo {author} {\bibfnamefont {M.}~\bibnamefont {Daira}}, \
	and\ \bibinfo {author} {\bibfnamefont {K.}~\bibnamefont {Yonemitsu}},\ }\href
{\doibase 10.1103/PhysRevB.97.115105} {\bibfield  {journal} {\bibinfo
		{journal} {Phys. Rev. B}\ }\textbf {\bibinfo {volume} {97}},\ \bibinfo
	{pages} {115105} (\bibinfo {year} {2018})}\BibitemShut {NoStop}%
\bibitem [{\citenamefont {Eckstein}\ and\ \citenamefont
	{Kollar}(2008)}]{Eckstein2008}%
\BibitemOpen
\bibfield  {author} {\bibinfo {author} {\bibfnamefont {M.}~\bibnamefont
		{Eckstein}}\ and\ \bibinfo {author} {\bibfnamefont {M.}~\bibnamefont
		{Kollar}},\ }\href {\doibase 10.1103/PhysRevB.78.205119} {\bibfield
	{journal} {\bibinfo  {journal} {Phys. Rev. B}\ }\textbf {\bibinfo
		{volume} {78}},\ \bibinfo {pages} {205119} (\bibinfo {year}
	{2008})}\BibitemShut {NoStop}%
\bibitem [{\citenamefont {Pedersen}\ \emph {et~al.}(2001)\citenamefont
	{Pedersen}, \citenamefont {Pedersen},\ and\ \citenamefont {{Brun
			Kriestensen}}}]{Pedersen2001}%
\BibitemOpen
\bibfield  {author} {\bibinfo {author} {\bibfnamefont {T.~G.}\ \bibnamefont
		{Pedersen}}, \bibinfo {author} {\bibfnamefont {K.}~\bibnamefont {Pedersen}},
	\ and\ \bibinfo {author} {\bibfnamefont {T.}~\bibnamefont {{Brun
				Kriestensen}}},\ }\href {\doibase 10.1103/PhysRevB.63.201101} {\bibfield
	{journal} {\bibinfo  {journal} {Phys. Rev. B}\ }\textbf {\bibinfo
		{volume} {63}},\ \bibinfo {pages} {201101} (\bibinfo {year}
	{2001})}\BibitemShut {NoStop}%
\bibitem [{\citenamefont {Kaneko}\ \emph {et~al.}(2012)\citenamefont {Kaneko},
	\citenamefont {Seki},\ and\ \citenamefont {Ohta}}]{Kaneko2012}%
\BibitemOpen
\bibfield  {author} {\bibinfo {author} {\bibfnamefont {T.}~\bibnamefont
		{Kaneko}}, \bibinfo {author} {\bibfnamefont {K.}~\bibnamefont {Seki}}, \ and\
	\bibinfo {author} {\bibfnamefont {Y.}~\bibnamefont {Ohta}},\ }\href {\doibase
	10.1103/PhysRevB.85.165135} {\bibfield  {journal} {\bibinfo  {journal}
		{Phys. Rev. B}\ }\textbf {\bibinfo {volume} {85}},\ \bibinfo {pages}
	{165135} (\bibinfo {year} {2012})}\BibitemShut {NoStop}%

\end{thebibliography}
%

\end{document}